\documentclass[twocolumn]{aastex63}
\usepackage{inputenc}
\usepackage{comment}
\usepackage{colortbl}

\received{10 Aug 2021}
\revised{17 Sept 2021}
\accepted{7 Oct 2021}
\submitjournal{ApJ}

\shorttitle{4FGL Spectra \& NNC}
\shortauthors{Kerby et al.}
\graphicspath{{./}{figures/}}

\begin{document}

\title{Multiwavelength Spectral Analysis and Neural Network Classification of Counterparts to 4FGL Unassociated Sources}

\author[0000-0003-2633-2196]{Stephen Kerby}
\affiliation{Department of Astronomy and Astrophysics \\
 Pennsylvania State University,
University Park, PA 16802, USA}

\author[0000-0002-0878-1193]{Amanpreet Kaur}
\affiliation{Department of Astronomy and Astrophysics \\
 Pennsylvania State University,
University Park, PA 16802, USA}

\author[0000-0002-5068-7344]{Abraham D. Falcone}
\affiliation{Department of Astronomy and Astrophysics \\
 Pennsylvania State University,
University Park, PA 16802, USA}

\author[0000-0003-0854-8313]{Ryan Eskenasy}
\affiliation{Department of Astronomy and Astrophysics \\
 Pennsylvania State University,
University Park, PA 16802, USA}
    
\author[0000-0002-0066-5083]{Fredric Hancock}
\affiliation{Department of Astronomy and Astrophysics \\
 Pennsylvania State University,
University Park, PA 16802, USA}

\author[0000-0002-3019-4577]{Michael C. Stroh}
\affiliation{Center for Interdisciplinary Exploration and Research in Astrophysics (CIERA) and Department of Physics and Astronomy, Northwestern University, Evanston, IL 60208, USA}

\author[0000-0001-7828-7708]{Elizabeth C. Ferrara}
\affil{NASA Goddard Space Flight Center, Greenbelt, MD 20771, USA}
\affil{Center for Research and Exploration in Space Science and Technology, NASA/GSFC, Greenbelt, MD 20771}
\affil{Department of Astronomy, University of Maryland, College Park, MD 20742}

\author[0000-0002-5297-5278]{Paul S. Ray}
\affiliation{Space Science Division, U.S. Naval Research Laboratory, Washington, DC 20375, USA}

\author{Jamie A. Kennea}
\affiliation{Department of Astronomy and Astrophysics \\
 Pennsylvania State University,
University Park, PA 16802, USA}

\author{Eric Grove}
\affil{Space Science Division, U.S. Naval Research Laboratory, Washington, DC 20375, USA}

\begin{abstract}

The \textit{Fermi}-LAT unassociated sources represent some of the most enigmatic gamma-ray sources in the sky. Observations with the \textit{Swift}-XRT and -UVOT telescopes have identified hundreds of likely X-ray and UV/optical counterparts in the uncertainty ellipses of the unassociated sources. In this work we present spectral fitting results for 205 possible X-ray/UV/optical counterparts to 4FGL unassociated targets. Assuming that the unassociated sources contain mostly pulsars and blazars, we develop a neural network classifier approach that applies gamma-ray, X-ray, and UV/optical spectral parameters to yield descriptive classification of unassociated spectra into pulsars and blazars.  From our primary sample of 174 \textit{Fermi} sources with a single X-ray/UV/optical counterpart, we present 132 $P_{bzr} > 0.99$ likely blazars and 14 $P_{bzr} < 0.01$ likely pulsars, with 28 remaining ambiguous. These subsets of the unassociated sources suggest a systematic expansion to catalogs of gamma-ray pulsars and blazars.  Compared to previous classification approaches our neural network classifier achieves significantly higher validation accuracy and returns more bifurcated $P_{bzr}$ values, suggesting that multiwavelength analysis is a valuable tool for confident classification of \textit{Fermi} unassociated sources.

\end{abstract}


\section{Introduction}
\label{sec:Intro}

The \textit{Fermi} Gamma-ray Space Telescope - Large Area Telescope (\textit{Fermi}-LAT) 4FGL catalog includes 5064 sources, 3376 of which are associated with extragalactic blazars or nearby pulsars \citep{Abdollahi2020}. 352 other 4FGL sources include supernova remnants, X-ray binaries, starburst galaxies, and other objects.  1336 4FGL sources are ``unassociated", lacking confident astrophysical explanations or source counterparts at longer wavelengths. Given that the bulk of the 4FGL sources are blazars or pulsars it is feasible that many of the unassociated sources are also blazars or pulsars that failed to be classified.   The unassociated sources therefore tease systematic expansions to blazar and pulsar catalogs by extending to less obvious sources.

Identifying blazars and pulsars among the unassociated 4FGL sources is an important step towards confident population studies of both classes.  The \textit{Fermi}-LAT unassociated sources might include blazars that are lower luminosity, higher redshift, or viewed at a larger angle with respect to the jet than their more easily detected and identified cousins in the established \textit{Fermi} blazar catalog \citep{Ferrara2015}.
A more complete blazar catalog will serve numerous scientific goals, including the verification and analysis of the blazar sequence \citep[e.g.,][]{Fossati1998,Ghisellini2017} as a theoretical unifying scheme for blazars.

Similarly, there may be undiscovered pulsars among the unassociated 4FGL sources, valuable additions to the significantly shorter list of \textit{Fermi} pulsars. The \textit{Fermi} pulsar list includes canonical and millisecond pulsars but is also notable for dramatically expanding research into `Black Widow' pulsars \citep[e.g.][]{Wu2018,KwanLok2018}, and the unassociated sources may contain several pulsars of these various types. Finally, some unassociated objects may defy classification as blazars or pulsars even after multiwavelength analysis.  Identifying and classifying the ``low-hanging fruit'' of pulsars and blazars among the unassociated sources is a first step in identifying and studying other astronomical objects.

While gamma-ray observations from \textit{Fermi}-LAT are the foundation for the 4FGL catalog, more confident classification of pulsars and blazars can be achieved by extending spectral analysis to lower energies. To this end, the Neil Gehrels \textit{Swift} Observatory (aka \textit{Swift}) \citep{Gehrels2004} is conducting a continuing survey of the \textit{Fermi} unassociated sources. With the XRT \citep{Burrows2005} and UVOT \citep{Roming2005} instruments onboard \textit{Swift}, X-ray observations from $0.3 - 10.0 \:\rm{keV}$ and UV-visual observations from $450 - 900 \:\rm{nm}$ systematically cover the uncertainty ellipse of each \textit{Fermi}-LAT unassociated source. 

Given the variety of spectral features in blazars and pulsars, this multiwavelength capability is a powerful tool for studying 4FGL unassociated sources at lower energies and for supporting identification and classification efforts. For example \cite{Kaur2019} sorted 217 high-S/N unassociated 3FGL sources into blazars and pulsars using machine learning on gamma- and a single X-ray spectral parameter. The sample was limited to only those sources with a single possible X-ray counterpart in the error ellipse. Training an ML routine with known gamma-ray pulsars and blazars, the authors identified $173$ likely blazars with $P_{bzr} >90 \%$ ($134$ with $P_{bzr} >99 \%$) and $13$ likely pulsars with $P_{bzr} <10 \%$ ($7$ with $P_{bzr} < 1 \%$). From their initial list, $31$ sources from the 3FGL unassociated list defied categorization and were labeled 'ambiguous'. Recent work by this collaboration \citep{Kerby2021} continued that analysis by including more detailed X-ray spectral fitting and systematically adding spectral parameters to machine learning. Adding UV-visual observations from the \textit{Swift}-UVOT telescope is particularly useful, as pulsars are usually extremely dim in the UV-visual range \citep{KwanLok2018, SazParkinson2016} while blazars emit at all wavelengths and at low redshift can be observed across the electromagnetic spectrum \citep{Ghisellini2008}.

The basis of expanding analysis of the unassociated sources to lower energies is the assumption that a gamma-ray source is likely to also be emitting X-rays due to the ubiquity of X-ray synchrotron emission in energetic gamma-ray systems. Given the spatial resolution and X-ray sensitivity of our observations with \textit{Swift}-XRT and the ubiquity of pulsars and blazars emitting both in gamma- and X-rays, if only a single X-ray source is present in a gamma-ray ellipse then there is likely a relationship between the two. Because gamma-ray emitters like pulsars and blazars normally emit X-rays via mechanisms like synchrotron radiation, an association between an unassociated gamma-ray source and a solitary X-ray sources within its uncertainty ellipse is theoretically sound. The automated analysis after \textit{Swift} observations calculates the probability of a coincident but unrelated X-ray source in the \textit{Fermi} 4FGL 95\% confidence region based on exposure time and confidence region size, and for a typical 4ks exposure in these $\sim$5 arcmin semi-major axis regions, the probability is $< 0.01$. For the handful of exposures with longer than 4ks and/or larger than typical 4FGL confidence regions sizes, the probability of additional spurious X-ray source detections increases.  Finally, there are a large number of 4FGL target fields for which no X-ray detection was found in the typical ~4 ksec exposure, which is consistent with the estimated low chance probability for spurious X-ray source detection.

In previous work on the \textit{Fermi} unassociated sources \citep[e.g.][]{Kaur2019, Kerby2021} analysis was restricted to \textit{Fermi} unassociated sources with only one high-S/N X-ray source in their gamma-ray confidence ellipse. Still, it may be the case that \textit{Swift} observations reveal multiple X-ray sources in the uncertainty ellipse of a single \textit{Fermi} unassociated gamma-ray source, as a 4FGL gamma-ray uncertainty ellipses spans several arcminutes while the positional uncertainty of a XRT detections is a few arcseconds.  In this case, other factors must be considered before determining a likely association between high- and low-energy photons.  It is less likely that there will be degeneracy between UV-visual and X-ray sources, as both the XRT and UVOT instruments on \textit{Swift} have relatively small PSFs.

Our approach to classification of 4FGL unassociated sources has several unique factors compared to other classification efforts and builds on our work on the 3FGL catalog. While \cite{Lefaucheur2017} conducted IR observations of unassociated sources to look for low-energy counterparts, skipping over X-ray/UV/optical observations would complicate attempts to use spectral information to link gamma-ray sources to counterparts. The sky is much more prolifically filled with IR sources than X-ray sources, and a gamma-ray uncertainty ellipse with just one possible X-ray counterpart might have dozens in the IR bands. Searching in the radio band is similarly difficult, requiring assumptions to link gamma-ray and radio emission without interpreting the intermediate wavelengths, but the detection of characteristic radio pulsations is a direct route to locating pulsars and some previous works have used radio properties to predict pulsar membership after efficient searches \citep[for example,][]{Frail2018}. Recently, \cite{Zhu2021} conducted ML on the 4FGL unassociated sources, but restricted their analysis to only gamma-ray properties.  Analysis of X-ray observations specifically can capture the synchrotron peaks of non-thermal emitters like blazars, a valuable region for discriminating the spectra of pulsars and blazars.

In this work, we extend investigations of the unassociated sources by combining gamma-ray data with \textit{Swift}-XRT and -UVOT observations for 4FGL unassociated targets observed thus far by \textit{Swift}'s systematic program. With this multiwavelength set of parameters, we build a neural network classification (NNC) routine using samples of known pulsars and blazars to classify the unassociated sources. In section \ref{sec:Samples}, we describe the gamma-ray, X-ray, and UV/optical observations and sources used in this work, plus reasons for excluding certain entries from our base sample.  Next, in section \ref{sec:Analysis} we describe our production of spectral parameters and present tabulated results.  In section \ref{sec:NNC} we discuss the classifications of the unassociated gamma-ray sources via a NNC method. In section \ref{sec:Results} we present and discuss spectral fitting and classification results, summarize our findings, and posit next steps.

\section{Observational Program, Targets, and Samples} 
\label{sec:Samples}

\subsection{\textit{Fermi}-LAT Unassociated and \textit{Swift}-XRT Sources}

The unassociated sources of the \textit{Fermi}-LAT 4FGL catalog are comprised of gamma-ray sources of unknown astrophysical nature and no known counterpart \citep[for a discussion of the entire 4FGL catalog, see][]{collaboration2019}. After ten continuous years of observations, the 4FGL-DR2 catalog \citep{Ballet2020} contains 5064 sources in total, of which over 1300 are unassociated. 1410 4FGL sources are targets for the \textit{Swift}-XRT survey of unassociated sources (slightly higher than the total number of currently unassociated sources due to sources gaining or losing associations between DR1 and DR2). By observing the unassociated gamma-ray sources, the \textit{Swift} program provides high-resolution X-ray, UV, and visual observations to find lower-energy counterparts to the unassociated sources. This survey is a further development after previously analyzing unassociated 3FGL sources \citep[detailed in][]{Kerby2021} and has covered approximately 500 4FGL unassociated targets with $> 4 \:\rm{ks}$ of observations each. So far \textit{Swift} has detected possible X-ray counterparts within the \textit{Fermi}-LAT uncertainty ellipse of 208 4FGL unassociated sources, with some unassociated sources having multiple possible X-ray counterparts.

Combining archival and new \textit{Swift} observations, an automated analysis process described in \cite{Falcone2011} produces lists of X-ray sources. For our purposes, a source is deemed `notable' if it is contained within the 95\% uncertainty ellipse of the 4FGL source and if the signal-to-noise ratio calculated with the \verb|Ximage| function \verb|SOSTA| is greater than $4$.  We take the produced list of 238 notable X-ray sources as containing the possible X-ray counterparts to the 4FGL unassociated sources, and focus on these detections as the sample for further X-ray analysis. 

The HEASARC query interface allows for downloads of \textit{Swift}-XRT observations within 8' of all 238 notable X-ray sources spread across the 205 4FGL targets.  Some X-ray sources were matched with \textit{Swift}-XRT observations only upon expanding the HEASARC search radius to 10', as they sit close to the edge of the field of view. Expanding the search radius of the HEASARC query does not impinge too closely towards the edges of the $23.6 \arcmin$ field of view of the \textit{Swift}-XRT telescope.

We immediately eliminate two 4FGL targets from consideration, along with all related possible counterparts. The two targets, J1836.8-2354 and J1649.2-4513, have very uneven \textit{Swift}-XRT coverage near the 4FGL centroid position due to a bright, highly-observed X-ray source just outside the field of view.  The nine notable X-ray sources from those two targets appear to be related to the extremely long ($> 100 \:\rm{ks}$) observations of the nearby X-ray source that is outside the uncertainty ellipse. Several other 4FGL targets are eliminated if their only possible X-ray counterpart is coincident with a catalogued star that contributes a false X-ray signal via optical loading, the pile-up of optical photons in the \textit{Swift}-XRT detector, or by a star of such brightness that it would radically dominate the UV/optical emission from an unassociated source, putting it outside of the range of known pulsars and blazars.

Of the 205 4FGL targets with possible X-ray counterparts examined in this work, 14 have more than one notable X-ray source within their uncertainty ellipse. For the purposes of spectral analysis and ML classification, we fit all X-ray sources while making no judgements about which is most likely the counterpart to the gamma-ray source. We tabulate and discuss the unitary sources as our primary sample separately from the more ``confused" unassociated sources with more than one X-ray/UV/optical possible counterpart.

\subsection{\textit{Swift}-UVOT Source Analysis}

For each X-ray source, we also used \textit{Swift}-UVOT data within 8-10$\arcmin$ of the XRT centroid to search for X-ray/UV/optical counterparts using the \textit{Swift}-UVOT telescope. For each XRT/UVOT source, we generated a fully integrated UVOT image by performing \verb|UVOTIMSUM| on the UVOT FITS file. We placed circular extraction regions at the coordinates of these UVOT detections; if there was no UVOT detection within the $\sim 5 \arcsec$ positional uncertainty of the XRT detection, we centered the extraction region on the X-ray source’s centroid position to determine an upper limit to the brightness. The radius of the UVOT extraction region was set to $5\arcsec$, the PSF of the \textit{Swift}-UVOT telescope for a point source. \verb|UVOTSOURCE| was then used to gather counts from the source region, and the background was measured using a source-free circular background region of radius $20 \arcsec$ elsewhere in the image for each position. The background-subtracted count rate was converted to flux and magnitude following the process described in \citet{breeveld2011updated} to obtain a UVOT magnitude in one of the bands in Table \ref{tab:UVOTbands}. UVOT counterparts imaged in multiple UVOT bands were analyzed once separately for each band.

We applied a $3\sigma$ detection threshold for UVOT counterparts using \verb|UVOTDETECT|. Though the vast majority of XRT detections had unique UVOT counterparts as well, fifteen regions had clustered UVOT sources that could not be disentangled within the constraints of the PSF of the telescopes on \textit{Swift}. Training, validation, and research datasets with complete column coverage are vital to the machine learning process described below, so we exclude those confused sources from the unassociated sample.

\section{Spectral Analysis}
\label{sec:Analysis}

\subsection{X-ray Spectral Analysis}

In total the HEASARC query for \textit{Swift}-XRT observations collected over $1000$ individual observations. Here, we only use observations in the photon counting (PC) mode of the \textit{Swift}-XRT, enabling two-dimensional imaging across the XRT field-of-view with reasonable energy resolution. Summed exposure time for the X-ray sources varies from a few kiloseconds to many dozens of kiloseconds.

Each level 1 event file was processed and cleaned using \verb|xrtpipeline| v.0.13.5 from the HEASOFT software\footnote{\url{https://heasarc.gsfc.nasa.gov/docs/software.html}}.  Only events graded 0 through 12 were used in this analysis. Merging the events and exposure files with other observations of each particular X-ray source was conducted using \verb|xselect| v.2.4g and \verb|ximage| v.4.5.1, resulting in a single summed event list for each source plus a summed exposure map and ancillary response file using \verb|xrtmkarf|.  This merging precludes any time series or variability analysis of the X-ray sources, but few X-ray sources in our sample have sufficient observation time or photon counts to enable detailed temporal analysis. 

For each possible X-ray counterpart, \verb|xselect| produced spectra for source and background regions. The source region was circular with radius 20 arcseconds, and the background region was annular with inner and outer radii of 50 and 150 arcseconds respectively. Both regions were centered on the coordinates of the examined X-ray source.  The background region size was chosen to be far outside the PSF of a point source for the \textit{Swift}-XRT detector (half-power diameter of $18 \arcsec$).

If the count rate in the center of the source region exceeded 0.5 counts per second (with PC mode having frames of $2.5 \:\rm{s}$ each, this is approximately greater than one photon per frame), drawing a new annular source region with an inner radius depending on the count rate would avoid photon pile-up and saturation on the detector. The possible X-ray counterparts to 4FGL unassociated sources are faint enough that none caused such pile-up. 

We used \verb|Xspec| v.12.10.1f \citep{Arnaud1996} to fit each spectrum. The fitting model included three nested functions: \verb|tbabs|, \verb|cflux|, and \verb|powerlaw|. \verb|cflux| calculated the total unabsorbed flux between 0.3 and 10 keV and \verb|tbabs| modeled line-of-sight hydrogen absorption using galactic values from the \verb|nH| lookup function described in \cite{Wilms2000}.  The galactic line-of-sight extinction is fixed at the catalog value for each spectrum analyzed. \verb|powerlaw| is a simple power law with photon index $\Gamma_X$. Uncertainties on the fitted photon index and X-ray flux were jointly measured using the iterative \verb|steppar| routine, and the spectral fitting results are included in an accompanying machine-readable database.

Fitting was executed using the C-statistic as the optimization metric. The C-statistic is a useful fitting statistic for spectra with few counts, particularly in cases for which there are not sufficient photons to bin counts for a $\chi^2$ fit \citep{Cash1976}. Compared to the $\chi^2$ statistic, which assumes Gaussian behavior in bins, the C-statistic operates under Poisson statistics. In this way, it is much more applicable to fitting spectra with very few counts in each energy bin. The C-statistic is given by

$$C = 2 \sum \left( (tm_i) - S_i +S_i ( \ln S_i - \ln (tm_i)) \right)$$

\noindent with $t$ the exposure time, $m_i$ the predicted count rate in any particular bin, and $S_i$ is the observed counts in each bin. For an X-ray source with only a few dozen detected X-ray photons, binning the events for a $\chi^2$ approach would result in only one or two bins, which is not useful for detailed spectral fitting. The Cash statistic does not require any such binning by assuming the more appropriate Poisson distribution of events.

The model incorporating hydrogen-absorbed power law spectra returned unusually high or low $\Gamma_X$ for nine 4FGL X-ray counterparts compared to the expected $0<\Gamma_X<4$ for pulsars and blazars. Given that some of these X-ray sources also coincided with catalogued stars, we viewed these sources as dubious for pulsar/blazar classification but worthy of additional investigation with other approaches. We do not include these spectra in the ML classification. These sources are listed in Table \ref{tab:trouble}.

\begin{deluxetable*}{ccrl}
\tablecaption{X-ray counterparts to 4FGL unassociated sources with extreme X-ray photon indices, $\Gamma_X < -1$ or $\Gamma_X > 5$, excluded from the main ML classification effort. Possible stellar counterparts include spectral type and apparent magnitude from SIMBAD if available.}
\label{tab:trouble}
\tablewidth{0pt}
\tablehead{
\colhead{4FGL Source} & \colhead{\textit{Swift} Source} & \colhead{$\Gamma_{X}$} & \colhead{Notes}}
\startdata
J0844.9$-$4117 & J084439.9$-$411642 & 6.51 & 6" from HD 74804, $m_V = 8.8$ young stellar object \\ 
J1214.7$-$5858 & J121400.6$-$585829 & 5.39 & \\
J1320.3$-$6410 & J132016.9$-$641353 & 8.19 & \\
J1639.8$-$4642 & J163946.2$-$464058 & 8.73 & 8" from TYC 8325-3618-1, $m_V = 10.5$ star\\
J1645.8$-$4533 & J164547.7$-$453647 & -1.53 & 7" from 4U 1642-45, low-mass X-ray binary\\
J1650.9$-$4420 & J165124.2$-$442141 & 10 & \\
J1706.5$-$4023 & J170633.3$-$402543 & 9.02 & 5" from Gaia young stellar object candidate \\
J1846.9$-$0227 & J184650.7$-$022903 & 7.90 & 4" from BD-02 4739, $m_V = 10.4$ star\\
J1902.2+0448 & J190220.7+044637 & 10 & 4" from BD+04 3956, $m_V = 10.1$ star \\
\enddata
\end{deluxetable*}

\subsection{V-magnitude Conversions}

The analysis of \textit{Swift}-UVOT data produces magnitudes in the six UVOT bands (vv, bb, uu, w1, m2, w2).  However, the training sample of known pulsars and blazars uses Johnson V magnitudes, requiring a conversion from the UVOT bands to the Johnson system.  To facilitate this conversion, we convert the UVOT magnitudes to fluxes, then use a power-law scaling relationship to predict the V-band flux, converting back to magnitude. The power-law scaling relation uses the central wavelengths of the UVOT bands (given in Table \ref{tab:UVOTbands}) and the V band ($540 \:\rm{nm}$) plus an assumed UV/optical spectral index of $\alpha_{c} = 0.5$.

\begin{deluxetable*}{c|cccccc}
\tablecaption{The central wavelengths of the six \textit{Swift}-UVOT bands used to convert \textit{Swift}-UVOT magnitudes to Johnson V magnitudes}
\label{tab:UVOTbands}
\tablewidth{0pt}
\tablehead{UVOT Band & vv & bb & uu & w1 & m2 & w2}
\startdata
Central Wavelength ($\rm{nm}$) & 547 & 439 & 347 & 260 & 225 & 193 \\
\enddata
\end{deluxetable*}

The ratio of two fluxes at two different wavelengths is simply

$$ \frac{F_1}{F_2} = \left( \frac{\nu_1}{\nu_2} \right)^{-\alpha_c} $$

\noindent which inserted into the magnitude equation gives a predicted conversion between the magnitude in UVOT band $m_i$ and in the Johnson V band.

$$ m_{i} - m_V = -2.5 \log{\frac{F_{i}}{F_V}} = 2.5 \alpha \log{\frac{\nu_i}{\nu_V}}  $$

Clearly, a conversion from the UVOT vv magnitude to the Johnson V magnitude is the most direct and most desirable conversion, because the UVOT vv band has close to the same central wavelength as the Johnson V band. Unfortunately, most \textit{Swift}-UVOT observations were in the higher-energy bands, requiring significant conversions and introducing uncertainties into the analysis. For \textit{Swift}-XRT sources with observations in multiple UVOT bands, we only used the converted V magnitude originating from the UVOT band closest to the Johnson V band, regarding that observation as most faithfully approximating the Johnson V magnitude.

\begin{figure}
    \centering
    \includegraphics[width=\columnwidth]{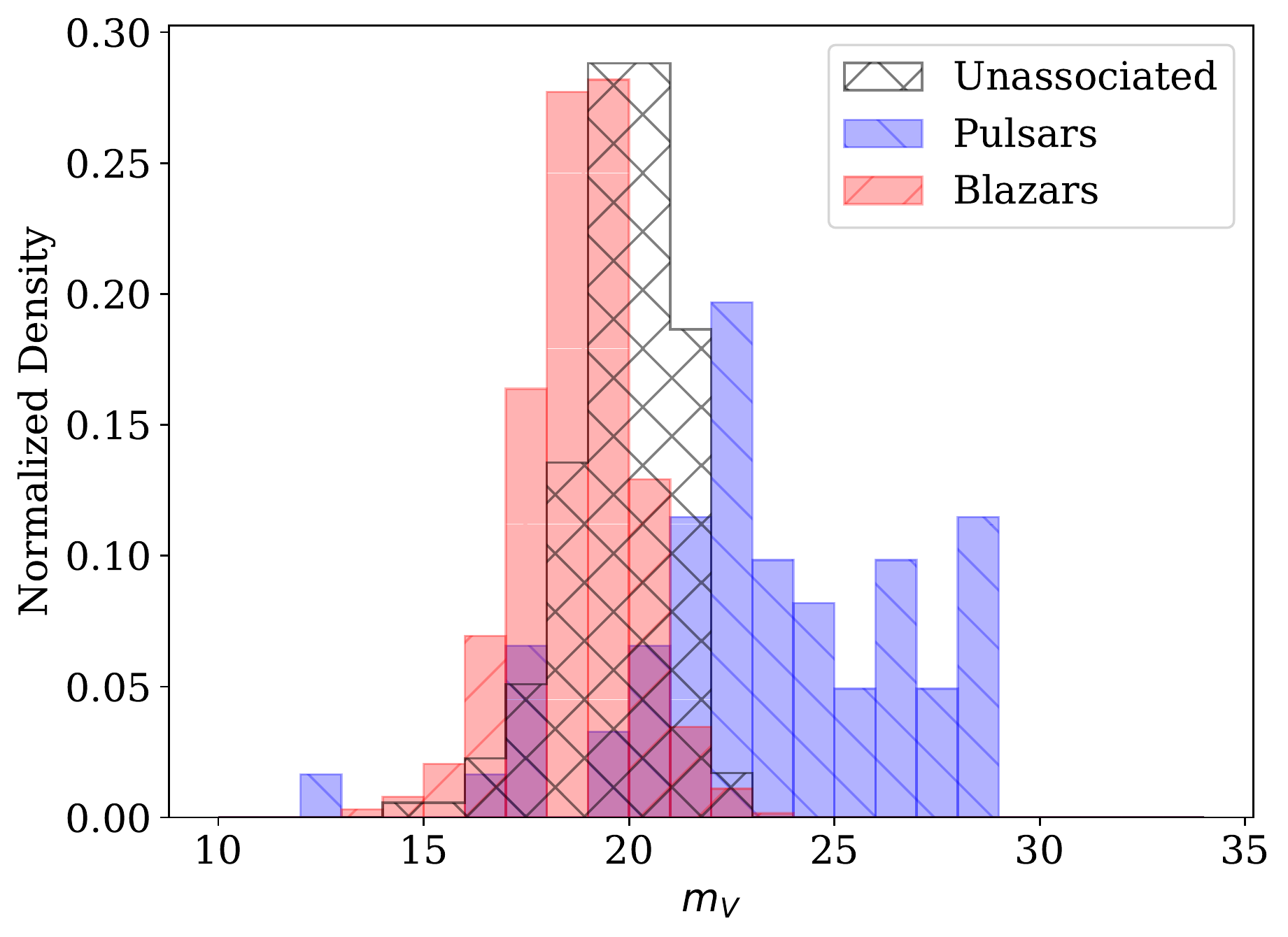}
    \caption{The converted V magnitude values for the known pulsar and blazar sample (in blue and red) and for the counterparts to the 4FGL sample identified with \textit{Swift}-UVOT (black lined). The difference in the distributions of known pulsars and blazars is great enough that the uncertainties in the magnitude conversion are small compared to the separation between the two distributions.}
    \label{fig:VmagHist}
\end{figure}

Conversion between one spectral band and another with an assumed slope is fraught with uncertainty. However, because the average V magnitudes of the known pulsar and blazar samples are so distinct (the medians separated by five magnitudes, at least a factor of 100x in flux), these uncertainties are tolerable as pulsars are typically much dimmer than blazars in visual and UV photons. The V magnitudes in our training and research samples are not corrected for extinction, which would mainly impact pulsars within the galactic plane. However, using the hydrogen column densities of the pulsars in our sample and the $n_H$ vs $A_V$ scaling relation given in \cite{Guver2009}, it is unlikely that an entire sample of pulsars is dimmed enough to create a fictitious separation in V magnitudes from the blazar population that would bias our classification efforts. Figure \ref{fig:VmagHist} shows the converted V magnitude distribution (using V magnitude upper limits for those pulsars without solid estimates) of the known \textit{Fermi} pulsars and blazars compared to the UV/optical counterparts in our unassociated sample.

\section{Neural Network Classification}
\label{sec:NNC}

\subsection{Known and Unknown Samples}
\label{sec:subSamples}

To train ML classification routines, we gathered a sample of 74 known gamma-ray pulsars (including radio-loud, radio-quiet, and millisecond pulsars) and 635 known gamma-ray blazars. This sample was derived from the catalog \textit{Fermi}-LAT pulsars and blazars \citep{Abdo2013,Ackermann2015}. The second \textit{Swift} X-ray Point Source Catalog \citep[2SXPS, ][]{Evans2020} provided X-ray parameters for many of the known pulsars and blazars, supplemented by a literature search of various studies \citep[e.g.,][]{Marelli2012,SazParkinson2016, Wu2018, Zyuzin2018}. We used the SIMBAD database to obtain V magnitudes as closely as possible, though for many sources magnitudes had to be converted from one optical band to the V band. 

Overall, our training sample is selected for those pulsars and blazars for which gamma-ray, X-ray, and optical brightness data is available, which implicitly limits the training of our classifier to that subsample of pulsars and blazars that have notable gamma-ray, X-ray, and optical flux.  This is especially impactful for pulsars, which have a wide range of intrinsic and extrinsic properties, including physical distance to our position in the galaxy. While we found that our training sample of pulsars is representative of catalogued gamma-ray pulsars in the \textit{Fermi}-LAT pulsar lists in terms of period and spin-down rate, canonical pulsars tend to be younger and more energetic than the general pulsar population. For pulsars, this leads to a training sample that preferentially includes nearby or energetic pulsars. Fortunately, our research sample of unassociated sources is similarly limited by X-ray and UV/optical flux, and has a similar fraction of subtypes of pulsars (65\% canonical, 35\% millisecond) compared to the overall \textit{Fermi}-LAT pulsar sample (55\% canonical, 45\% millisecond). Any sources deemed likely pulsars by our classification efforts are astrophysically similar to our training sample.

Preferring parameters that are distance-independent, we expressed the X-ray and V-band brightness in terms of a ratio with the gamma-ray flux. While $\log{F_X/F_\gamma}$ is a simple ratio of two fluxes\footnote{In this work, $\log{x}$ always refers to the logarithm in base 10 of $x$.}, the conversion of V magnitude to flux was slightly more complicated. Using the magnitude equation we converted the V magnitudes to fluxes using a reference flux and magnitude $m_2$ and $F_2$. However, because we are taking the logarithm of the ratio between V-band flux to gamma-ray flux, and because ML procedures first rescale and recenter each parameter, the exact reference flux and magnitude used herein are irrelevant. For our purposes, we adopt the conversion from \cite{Bessell1998}.

$$ \log{F_V} = -\frac{m_V + 21.1}{2.5} $$

The parameters for each training or research source include:
\begin{itemize}
    \item X-ray photon index, $\Gamma_X$
    \item Gamma-ray photon index, $\Gamma_\gamma$ (\verb|PL_Index| in the 4FGL catalog)
    \item The logarithm of gamma-ray flux, $\log{F_\gamma}$, in $\rm{erg/s/cm^2}$ (\verb|Energy_Flux| in the 4FGL catalog)
    \item The logarithm of X-ray to gamma-ray flux ratio, $\log{F_X/F_\gamma}$
    \item The logarithm of V-band to gamma-ray flux ratio, $\log{F_V/F_\gamma}$.
    \item The significance of the curvature in the gamma-ray spectrum, henceforth simply \textit{curvature} (\verb|PLEC_SigCurv| in the 4FGL catalog)
    \item The year-over-year gamma-ray variability index (\verb|Variability_Index| in the 4FGL catalog)
\end{itemize}

\begin{figure*}
    \centering
    \includegraphics[width=\linewidth]{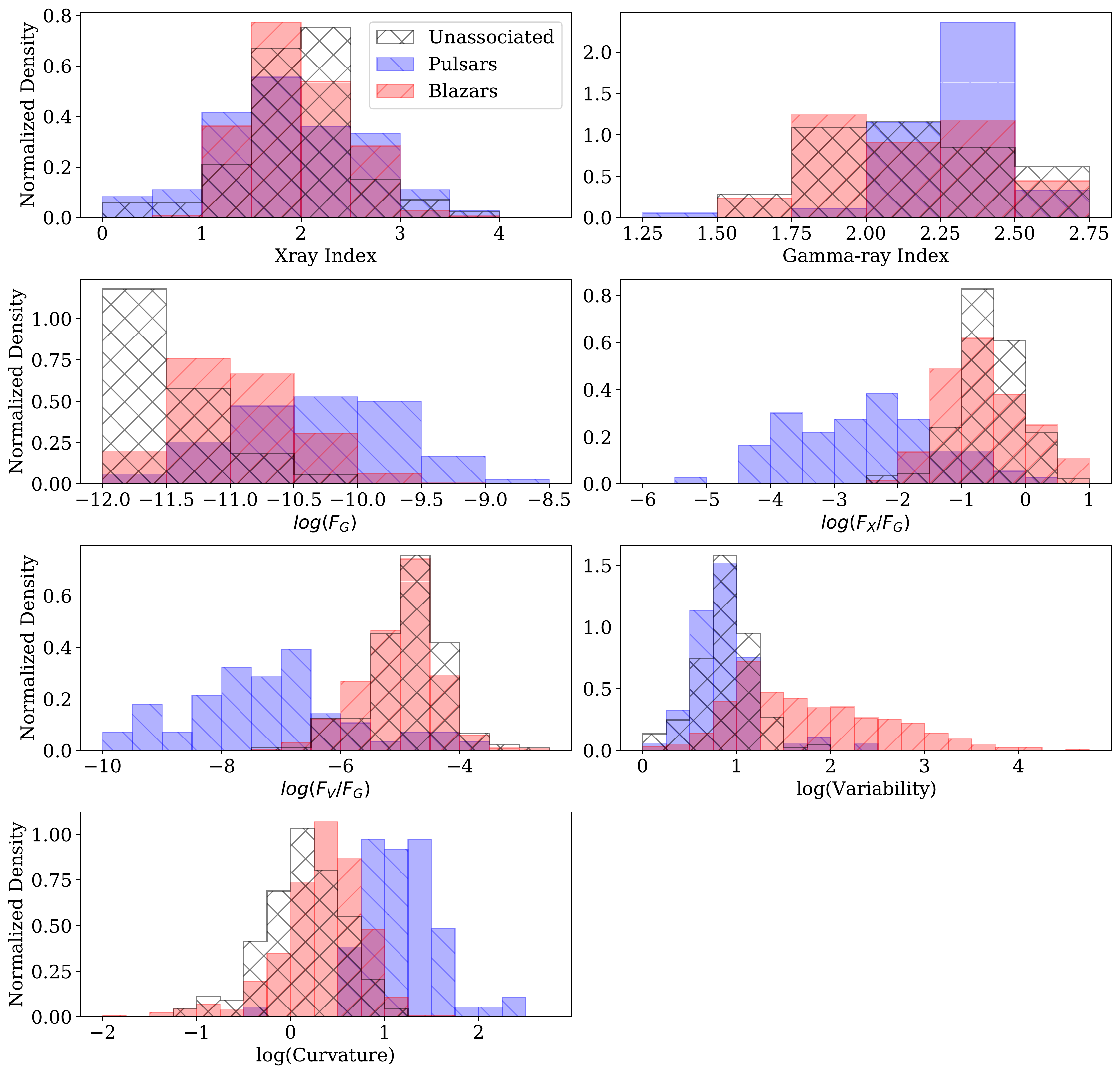}
    \caption{Normalized histograms for the known pulsars (blue), known blazars (red), and unassociated (black line) samples.}
    \label{fig:AllHistos}
\end{figure*}

Normalized histograms of all parameters for both the known pulsars and blazars and the unassociated spectra are shown in Figure \ref{fig:AllHistos}. For the 174 X-ray sources that are the unique XRT/UVOT possible counterpart to the respective unassociated targets in our research sample, the gamma-ray, X-ray, and UV/optical parameters are given in Table \ref{tab:xrt}. For the unassociated targets with multiple possible counterparts in the uncertainty ellipse, the parameters are given in Table \ref{tab:xrt2}.

It is worth noting that while the histograms in Figure \ref{fig:AllHistos} show notable overlaps between the unassociated and known samples, there are still certain qualitative differences between the samples.  The unassociated sources have systematically lower gamma-ray flux than the known blazars and pulsars, meaning that the variability and curvature estimates from the 4FGL catalog have weaker photon statistics. A bright \textit{Fermi}-LAT blazar with significant variability on many timescales might not have observable variability if moved to a greater distance for no other reason than the Poisson-distributed arrival statistics of the photons. Still, the current flux-limited samples of pulsars and blazars \citep[for example, the flux- and redshift-limited sample of \textit{Fermi} blazars in ][]{Ghisellini2017} suggests that there are some objects left out of the \textit{Fermi} association lists simply due to being slightly dimmer.

Because there are many more known blazars than known pulsars in the training sample, we used Synthetic Minority Over-sampling Technique (SMOTE) \citep{Chawla2002} to generate additional pulsars that mirror the distribution of real pulsars with a k-nearest neighbors approach.  Previous classification efforts have shown that unbalanced training datasets can lead to classifiers that are biased against the underrepresented class \citep[for example,][]{Last2017}. The result of the SMOTE expansion is a training set with an equal number of blazars and pulsars, with the artificial pulsars being generated via SMOTE using the real pulsar distribution. The final database is eventually split into training and validation subsamples which contain real known blazars, real known pulsars, and SMOTE-generated `known pulsars' that mirror the distributions of spectral properties of the real known pulsars.

\subsection{Neural Network Design and Training}

Building the NNC method, we constructed an approach where the unassociated sources will be assigned a blazar probability $P_{bzr}$ depending on their similarity with known pulsars and known blazars. $P_{bzr} = 0$ denotes a $0\%$ probability of the object being a blazar, while $P_{bzr} = 1$ corresponds to $100\%$ probability the object is a blazar. The NNC had seven input nodes (one for each parameter of the training dataset), one hidden layer with four neurons, and one output node returning the predicted blazar probability. In this research we use the \verb|MLPClassifier| function of the \verb|scikit-learn| Python package for NNC training and validation, training the NNC with the `adam' optimization approach \citep{Kingma2014}. To allow for validation and accuracy checks for the trained NNC, we took a random selection ($20\%$ of the training dataset) as a validation subsample, leaving the remainder of known pulsars and blazars as the training subsample. The random selection method \verb|StratifiedShuffleSplit| was used so that the training and validation subsamples would have exactly the same proportion of pulsars and blazars.

To determine when to stop iterating and training the NNC, we relied on the \textit{Log-Loss} parameter, an error measure in binary classification approaches that is similar to a likelihood estimator. For a sample with true classifications $y_i = (0,1)$ and predicted classification $p_i \in [0,1]$, the Log-Loss is given by

$$ L_{log} = - \sum_i \left( y_i \log(p_i) + (1-y_i) \log(1-p_i) \right)$$

\begin{figure}
    \centering
    \includegraphics[width=0.9\columnwidth]{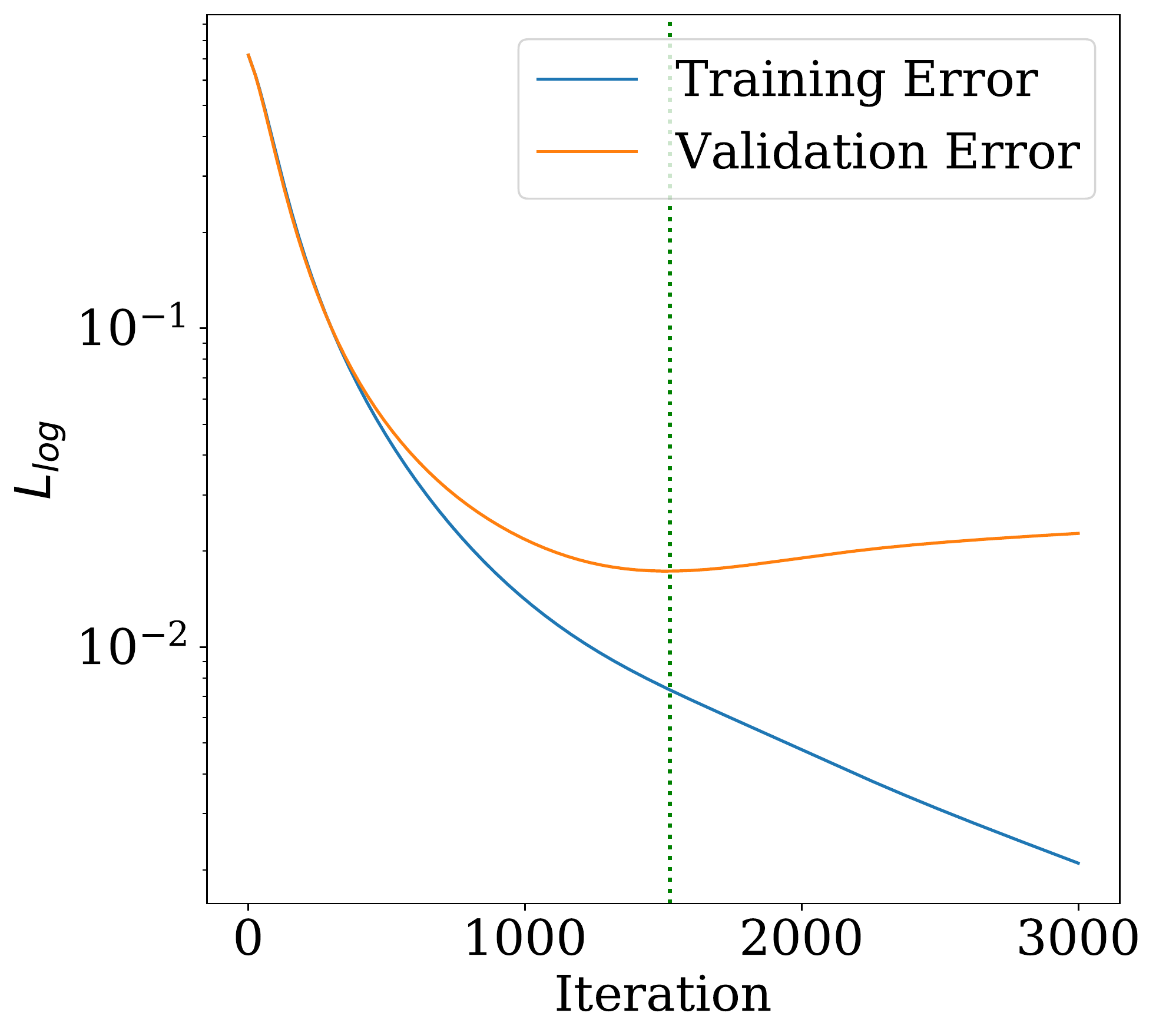}
    \caption{The evolution of the $L_{log}$ parameter as NNC iteration continues.  When the validation error levels off, iteration is stopped to prevent overfitting and loss of predictive capability.}
    \label{fig:LogLoss}
\end{figure}

At each iteration step in training the NNC, we calculate $L_{log}$ of both the training and validation subsamples.  As training continues, the training subsample $L_{log}$ should continuously decrease as the NNC approaches a more exacting fit. However, at some point the validation $L_{log}$ begins to increase as the NNC starts to overfit the training sample and lose predictive capabilities on unknown data points.  At this point, training is stopped. Figure \ref{fig:LogLoss} shows how the NNC training is stopped at approximately 2000 iterations when the validation $L_{log}$ levels out.

After using the training subsample to construct the NNC, we passed the validation subsample through the NNC and recorded the predicted blazar probabilities. Ideally, the NNC would predict blazar probabilities of $P_{bzr}=0$ or $P_{bzr}=1$ for the validation pulsars or blazars. A validation score for the NNC method depends on the cutoff used to determine what constitutes a ``likely" pulsar or plazar.  For example, a $90\%$ cutoff would designate any source with $P_{bzr} < 0.1$ a pulsar, with $P_{bzr} > 0.9$ a blazar, while a $99\%$ cutoff would only capture sources with $99\%$ confident classifications.

An optimally trained NNC should result in $P_{bzr}$ as close to $0$ or $1$ as possible for the validation subsample.  Typically, NNCs are judged based on the proportion of validation scores above some cutoff value; in a two-class example, the default cutoff for a `correct' classification is normally $0.5$. However, this single cutoff score does not investigate the degree of confidence with which the NNC is classifying the validation sources; a NNC which grades all validation pulsars at $P_{bzr} = 0.3$ and all validation blazars at $P_{bzr} = 0.7$ would have a $100\%$ validation score with a cutoff of $0.5$ but a $0\%$ score if one strives for greater confidence in classification.

\begin{figure}
    \centering
    \includegraphics[width=0.9\columnwidth]{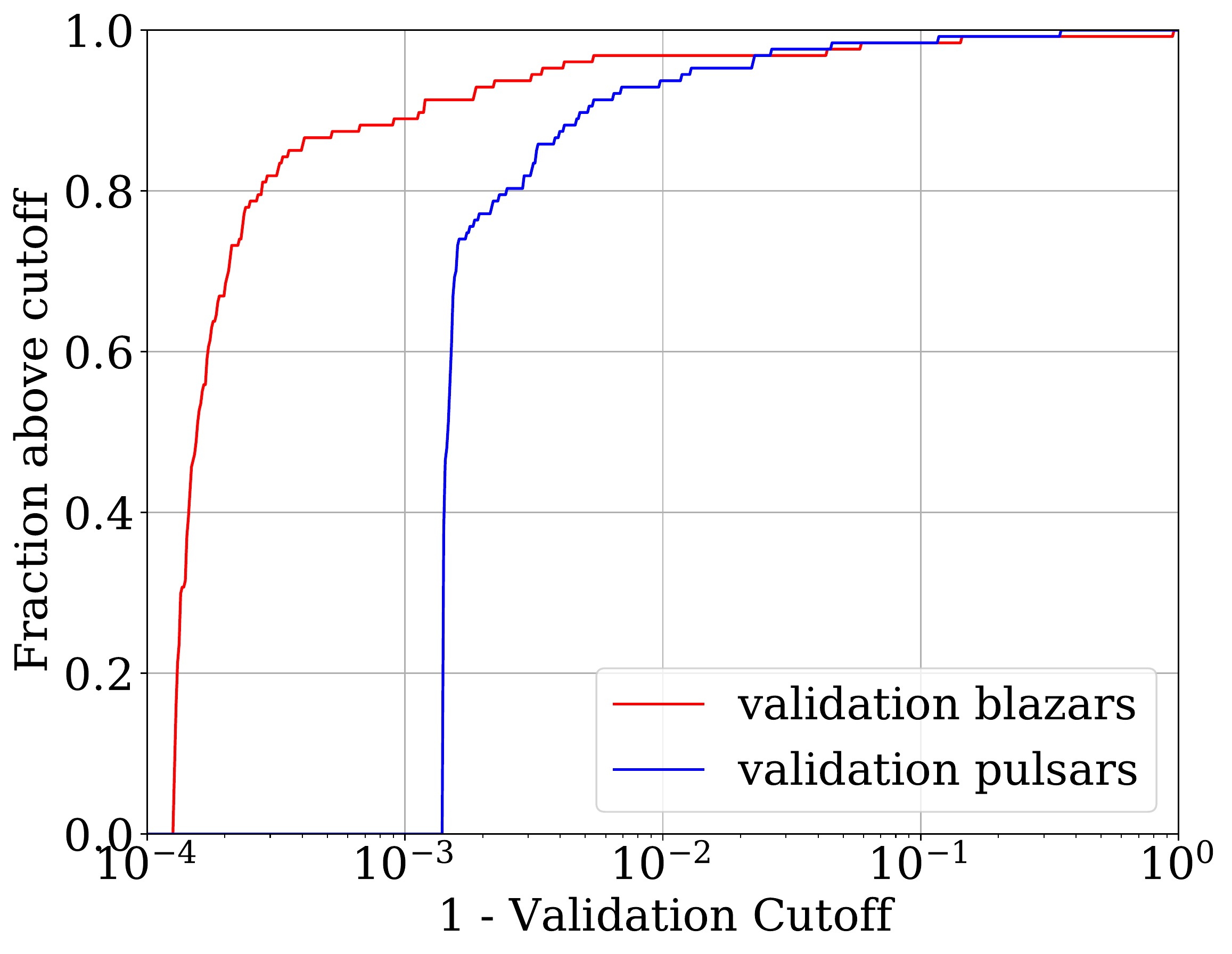}
    \caption{The fraction of validation pulsars (blue) and blazars (red) classified with $P_{bzr}$ above different cutoff values after training the NNC. The NNC classifies validation blazars with greater $P_{bzr}$ than validation pulsars, with no validation pulsars having $P_{bzr} < 10^{-3}$ but many validation blazars having $P_{bzr} > 0.999$}
    \label{fig:NNCcdp}
\end{figure}

Figure \ref{fig:NNCcdp} shows a generalized approach for determining both the reliability and confidence of classifications from a NNC.  Scanning through a logarithmic range of possible cutoff scores approaching unity (with $P_{psr} = 1-P_{bzr}$ for pulsars), the figure shows how the fraction of known pulsars and blazars correctly classified as such during the validation step changes depending on the $P_{bzr}$ cutoff used. This general approach illustrates progress classifying pulsars and blazars at higher degrees of confidence upon adding additional spectral parameters to the samples.

Applying the default $50\%$ cutoff used in previous classification efforts (i.e. a validation blazar must just have a validation $P_{bzr} > 0.5$ to be considered ``correctly" classified) returns an overall accuracy of $99.2\%$ for pulsars and blazars. However, a major improvement of this NNC approach to previous classification efforts \citep{Kaur2019,Kerby2021} is its capability to classify validation pulsars and blazars with over $99\%$ confidence. As shown in Figure \ref{fig:NNCcdp}, over $90\%$ of validation pulsars and blazars are classified with greater than $99\%$ confidence in the correct categories, buoying our hope that this NNC routine can pick likely blazars and pulsars from the unassociated sample.

\begin{figure}
    \centering
    \includegraphics[width=0.9\columnwidth]{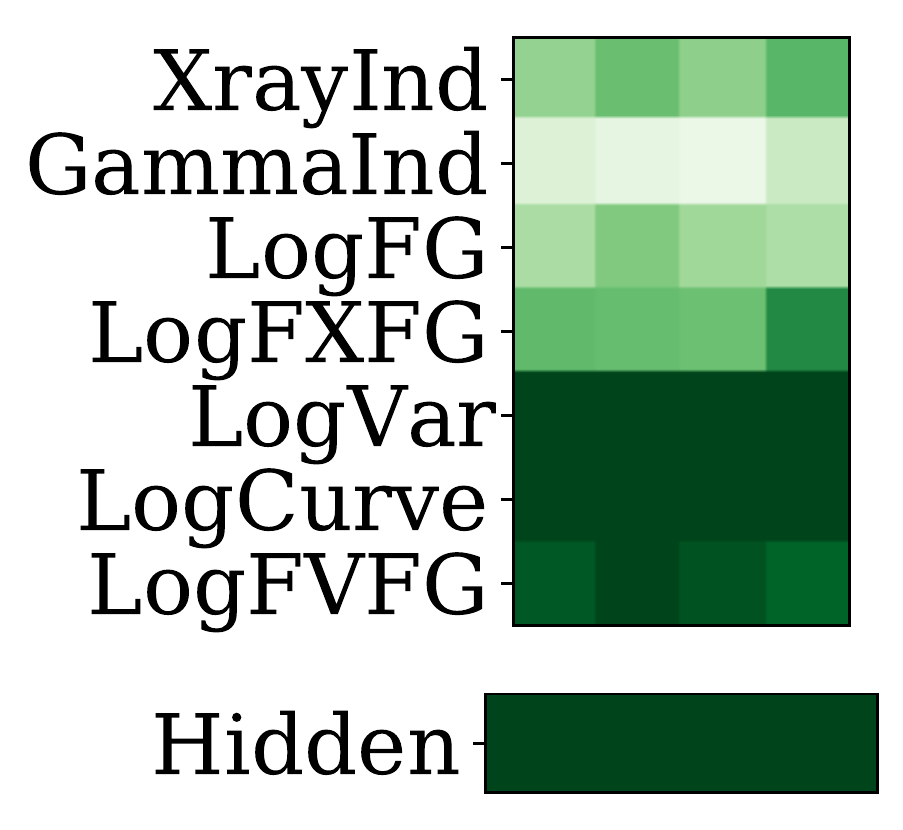}
    \caption{A qualitative visualization of the importances of the spectral features in our NNC routine. The columns represent the four neurons in our single hidden layer, the rows of the upper plot representing the seven spectral features used herein. Squares that are shaded darker have linearly heavier weights in the NNC, and therefore play a more important role in discriminating pulsars from blazars.}
    \label{fig:NNCImport}
\end{figure}

Calculating importances of the different parameters in a neural network ML approach is slightly more difficult than in other related classification schemes. Neurons in an NNC use an activation function which means that certain parameters may not be considered at all for certain entries. Still, the NNC is described by a weight matrix, and for a network with only a single hidden layer these weights are appropriate measures of the importance of different features. Figure \ref{fig:NNCImport} shows a visualization of the weights of the different parameters in our NNC approach, with darker shaded squares representing more impactful features in classifying blazars and pulsars.

\section{Results and Discussion}
\label{sec:Results}
\subsection{Unassociated Classification Results}

After training and validating the NNC, we recorded output $P_{bzr}$ values for each of spectra in the unassociated sample.  Of the 205 total fully examined sources, we found 157 with $P_{bzr} > 0.99$ and 18 with $P_{bzr} < 0.01$.  As some of the unassociated targets have more than one notable X-ray source in the gamma-ray uncertainty ellipse, additional work is needed to decipher the results around those targets including which if any X-ray possible counterpart is the actual partner to the gamma-ray emission. However, of the 174 X-ray sources that are the unique XRT/UVOT possible counterpart to the unassociated gamma-ray source, 14 have $P_{bzr}<0.01$ and 132 have $P_{bzr}>0.99$. These portions of our research dataset are highly likely pulsar and blazar candidates. The results of the NNC classification on the 4FGL sources with just a single possible X-ray counterpart are given in the last column of Table \ref{tab:xrt}, organized into likely pulsars ($P_{bzr} < 0.01$), likely blazars ($P_{bzr} > 0.99$) and ambiguous sources.

Several of the 4FGL targets examined herein have since been the subject of subsequent discoveries that shed additional light on their nature. \cite{Li2018} discusses the detection of a redback pulsar near 4FGL J0955.3-3949, which matches both for X-ray spectral parameters and in location on the sky of the X-ray counterpart in our analysis.  4FGL J0212.1+5321 has also been connected to a millisecond redback pulsar \citep{Linares2017,Li2016}. 4FGL J2039.5-5617 has recently been identified as a redback pulsar \citep{Clark2021}, as has 4FGL J1306-4035 after observations by the Parkes radio telescope \citep{Keane2018}, and 4FGL J1304.4+1203 is unassociated in 4FGL but classified as a pulsar in 4FGL-DR2 \citep{Ballet2020}. These sources are classified as likely pulsars in this work, showing that our NNC can independently predict classifications for sources that in other works are verified.  This increases our confidence that our classifier can point towards other promising sources to examine, and that the link between \textit{Swift} X-ray/UV/optical and \textit{Fermi}-LAT gamma-ray sources is valuable and well-founded. Additionally, 4FGL J0838.7-2827 and J0523.3-2527, sources that remained ambiguous in this work, have also since been identified as redback pulsars \citep{Halpern2017,Strader2014}.

To gain additional insight into 4FGL unassociated sources with multiple notable X-ray sources, we leveraged the much higher spatial resolution of the \textit{Swift} telescopes to do a coordinate search at the positions of X-ray counterparts. For many lower-energy counterparts, this search returned a coincident catalogued object and provided useful information to decide if the XRT/UVOT source is a likely counterpart to the unassociated target. Table \ref{tab:xrt2} includes the NNC classification results of these possible counterparts grouped by 4FGL target, and Table \ref{tab:xrtmultcross} lists the results of our spatial cross-reference. Several possible counterparts are coincident with catalogued stars, X-ray binaries, or galaxies, differences that might be useful for finding the most likely counterpart among multiple near a single 4FGL target. Interestingly, many of the pairs of possible counterparts have very similar $P_{bzr}$ values, suggesting that the 4FGL source in question may be a pulsar or blazar regardless of which X-ray source is truly linked with the gamma-ray emission.

The classification of the unassociated sample is more bifurcated in terms of $P_{bzr}$ compared to previous works \citep{Kaur2019,Kerby2021}; most of the unassociated sources have $P_{bzr}$ very close to $0$ or $1$, showing that the NNC makes confident predictions of blazar or pulsar class membership. The NNC's high validation accuracy and confidence suggest that the unassociated sources can be classified properly if they are pulsars and blazars similar to the known pulsars and blazars in our training sample. Even outside the context of the NNC as a whole, the histograms in Figure \ref{fig:AllHistos} and the feature weights in Figure \ref{fig:NNCImport} show that $\log{F_X/F_\gamma}$ and $\log{F_V/F_\gamma}$, flux ratios introduced in this work, are important parameters to distinguish pulsars from blazars, with pulsars having systematically lower values of both.

While previous approaches have classified unassociated sources with gamma-ray spectral parameters alone, Figure \ref{fig:AllHistos} shows that X-ray/UV/optical properties of counterparts to 4FGL targets can dramatically improve the discrimination of pulsars from blazars.  While the unassociated sources have lower gamma-ray flux than either of the known pulsar/blazar distributions, and the gamma-ray photon index is not a particularly useful discriminatory variable in any capacity, the variability and spectral curvature measures in the \textit{Fermi} catalog send mixed signals. The unassociated sources have the low variability expected of pulsars, but the low spectral curvature expected of blazars, probably due to their inherently lower photon statistics. Our addition of several new distant-independent properties via \textit{Swift}-XRT and -UVOT analysis adds three new features; Figure \ref{fig:NNCImport} shows that $\log{F_V/F_G}$ and $\log{F_X/F_G}$ are features of moderate to major importance in the NNC weight matrix, as is the X-ray photon index.  This reinforces the importance of systematic and methodical follow-up observations of unassociated gamma-ray sources at lower energies to uncover their astrophysical justification.

Overall, the spectral analysis and classification using \textit{Swift}-XRT and -UVOT data herein is valuable not only for characterizing the \textit{Fermi} unassociated sources, but also for guiding follow-up observations with likely pulsars and blazars for targeted searches while supporting numerous additions to catalogs of known gamma-ray pulsars and blazars.

\subsection{Comparison with Random Forest Classifier}

A random forest (RF) classifier uses an array of decision trees to classify an object into one of several categories. Each individual decision tree consists of several inequalities using the different parameters of a dataset in a ``choose your own adventure"-style series of judgements; compounded many times, a RF classifies an unknown object based on the fraction of the constituent trees giving each classification.

Previous works by this group on earlier catalogs of \textit{Fermi} unassociated sources \citep{Kaur2019,Kerby2021} have used RF classifiers to discern likely pulsars from likely blazars among the \textit{Fermi}-LAT unassociated sources.
While we have used an NNC approach in this work, it is worthwhile to compare the NNC $P_{bzr}$ output to the values produced by a RF classifier. Though the NNC and RF approaches make independent predictions for $P_{bzr}$ on the unassociated sources, the two approaches giving similar outputs on the same dataset would increase confidence in the reproducibility of our results even with different classification approaches. Additionally, comparing the two methods illuminates whether one is more descriptive or predictive in its $P_{bzr}$ output on our multiwavelength data. 

\begin{figure}
    \centering
    \includegraphics[width=\columnwidth]{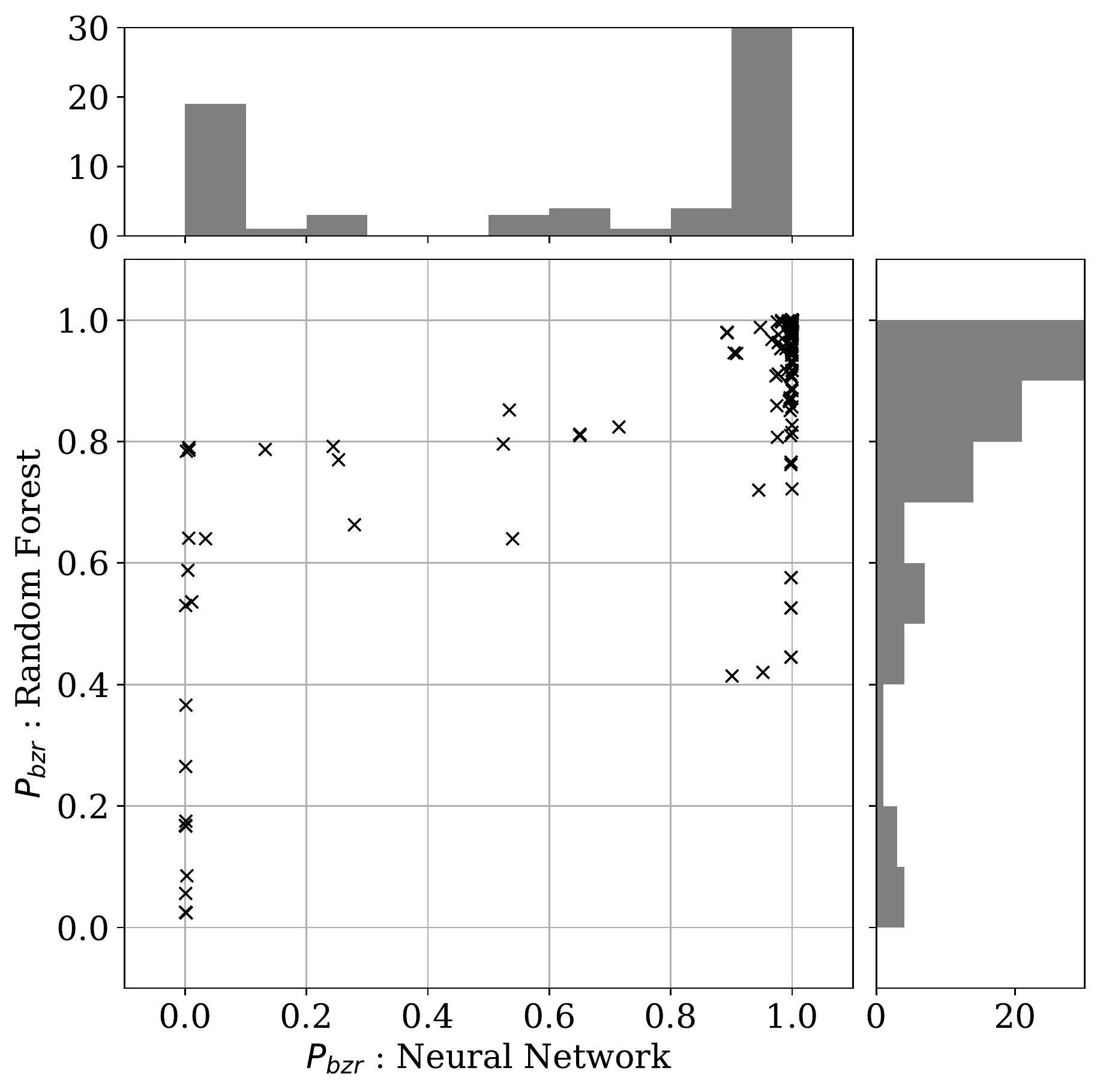}
    \caption{Comparing the $P_{bzr}$ scores for the 4FGL unassociated sample, using the exact same training and unknown datasets. The histograms of the two datasets have limited Y-axes to show the sparsely-populated bins between the two extremes. In both cases, the number of sources with $P_{bzr}$ close to unity is over 200.}
    \label{fig:RFvsNNC}
\end{figure}

We applied the RF classifier developed in \citet{Kerby2021} to the 4FGL sample of this paper. Figure \ref{fig:RFvsNNC} shows that while the NNC and RF both classify many sources with $P_{bzr}$ close to $0$ or $1$, the NNC classifies many sources as high or low $P_{bzr}$ that the RF leaves ambiguous. Indeed, the NNC results only have a few sources with $P_{bzr}$ not close to $0$ or $1$, and these sources almost uniformly have RF $P_{bzr}$ values around $0.8$.  This trend suggests that the NNC is more sensitive to pulsars than the RF approach, classifying sources as likely pulsars that in the RF method are ambiguous.  The more bifurcated nature of the NNC $P_{bzr}$ values compared to the RF results suggests that our NNC approach tends towards more confident classification and leaves fewer sources ambiguous, producing more immediately verifiable results, rather then hedging with ambiguity.

\subsection{Summary and Next Steps}

In this work we have classified 174 unique gamma-ray/X-ray/UV/optical sources into 14 likely pulsars, 132 likely blazars, and 28 ambiguous. Using \textit{Swift}-XRT and -UVOT observations, we built a collection of results, presented in Tables \ref{tab:xrt} and \ref{tab:xrt2}, representing a significant collection of observations within the uncertainty ellipse of the 4FGL unassociated gamma-ray sources. It is likely that the X-ray/UV/optical sources described herein have the same astrophysical origin as the gamma-rays described in the \textit{Fermi}-LAT unassociated catalog, so the unique correspondences in Table \ref{tab:xrt} describe the lower-energy spectra of the astronomical objects behind the gamma-ray emission of \textit{Fermi} unassociated sources.

Next, we built a neural network classification approach to use spectral information to divide the unassociated sample into likely blazars and pulsars using samples of known gamma-ray blazars and pulsars, reaching higher accuracy on validation subsamples and greater confidence in classification of unassociated sources than previous approaches. Of the 174 unique gamma-ray/X-ray/UV/optical spectra constructed and described in Table \ref{tab:xrt}, 132 are $P_{bzr} > 0.99$ likely blazars and 14 are $P_{bzr} < 0.01$ likely pulsars.  Leveraging the advantages of multiwavelength analysis, our new subsamples of likely pulsars and blazars can expand known gamma-ray pulsar and blazar catalogs to include sources with lower gamma-ray luminosity that were previously unassociated. 

As \textit{Swift} continues its observation campaign of the \textit{Fermi}-LAT unassociated sources, additional X-ray sources will be detected around previously unobserved targets. Planned follow-up observations across the electromagnetic spectrum can continue to investigate interesting sources discovered herein, including likely pulsars and ambiguously classified objects. These additional observations could validate likely pulsar classification with radio observations near the X-ray source to detect radio pulsations, or investigate ambiguous sources in greater detail to discern their true origin.

The subsamples of likely pulsars and blazars classified with our NNC approach are prime candidates for inclusion in population studies of gamma-ray pulsars and blazars. For example, using archival infrared observations of the likely blazars should allow for classification into likely BL Lac or likely FSRQ subsets, illuminating the biases and drawbacks of the \textit{Fermi} blazar catalogs used to investigate blazars as a class of AGN.  The likely pulsar subset, if validated with radio pulsation searches, would expand the list of known gamma-ray pulsars by almost 10\%.

\software{Astropy \citep{Astropy}, numpy \citep{NumPy}, Matplotlib \citep{Matplotlib}, scikitlearn \citep{Scikitlearn}, FTools \citep{FTOOLS}}

\acknowledgments

This research has made use of data and/or software provided by the High Energy Astrophysics Science Archive Research Center (HEASARC), which is a service of the Astrophysics Science Division at NASA/GSFC. We gratefully acknowledge the support of NASA grants 80NSSC17K0752 and 80NSSC18K1730. E. Ferrara is supported by NASA under award number 80GSFC17M0002.

\textit{Fermi} research at NRL is supported by NASA.

\clearpage

\bibliography{4FGL}{}

\begin{longrotatetable}

\begin{deluxetable*}{ccccccccccc}
\tablecaption{\textit{Fermi}-LAT features for the unassociated sample investigated in this work, along with \textit{Swift}-XRT and -UVOT parameters for likely X-ray/UV/optical counterpart. Only unassociated sources with a single possible counterpart are included in this table, and sources are organized based on $P_{bzr}$, listing likely pulsars and likely blazars separately. The extraction and derivation of the parameters here are described in sections \ref{sec:Analysis} and \ref{sec:Samples}. The $m_V$ estimates were produced using the noted UVOT filter, the closest available to the V band central wavelength. \label{tab:xrt}}

\tablewidth{\columnwidth}
\tablehead{
\colhead{Target} & \colhead{$\log(F_G)$} &\colhead{$\Gamma_G$} &
\colhead{Vari. Ind.} & \colhead{Curv. Ind.}&\colhead{XRT source}& \colhead{$\log(F_X)$} & \colhead{$\Gamma_X$} & \colhead{UVOT Filter} & \colhead{$m_V$} & \colhead{$P_{bzr}$} \\
\colhead{4FGL} & \colhead{$\log(\rm{erg/s/cm^2})$} &\colhead{} &
\colhead{} & \colhead{} & \colhead{SwXF4} & \colhead{$\log(\rm{erg/s/cm^2})$} & \colhead{} & \colhead{} & \colhead{} & \colhead{}
}
\startdata
 &  &  &  &  &  &  &  &  &  & \\
 &  &  &  &  & (a) \large{$P_{bzr} < 0.01 $} &  &  &  &  & \\
 &  &  &  &  &  \large{\textbf{likely pulsars}} &  &  &  &  & \\ \arrayrulecolor{gray} \hline
J0031.0-2327 & -11.72 & 2.19 & 5.56 & 2.89 & J003040.1-232821 & -12.45 & 0.40 & uu & 21.39 & 0.0080 \\ \hline 
J0212.1+5321 & -10.79 & 2.21 & 3.24 & 11.08 & J021210.7+532136 & -11.81 & 1.09 & uu & 16.14 & 0.0053 \\ \hline
J0859.3-4342 & -10.61 & 2.64 & 3.27 & 5.90 & J085926.2-434526 & -11.65 & 1.05 & uu & 20.55 & 0.0029 \\ \hline 
J0955.3-3949 & -10.89 & 2.45 & 9.28 & 6.58 & J095527.3-394747 & -12.42 & 0.90 & uu & 21.49 & 0.0028 \\ \hline 
J1120.0-2204 & -10.79 & 2.17 & 5.24 & 14.26 & J111959.2-220456 & -13.11 & 1.86 & vv & 19.99 & 0.0018 \\ \hline 
J1304.4+1203 & -11.70 & 2.27 & 3.56 & 5.88 & J130442.8+120440 & -12.91 & 1.43 & uu & 21.43 & 0.0028 \\ \hline 
J1306.8-4035 & -11.07 & 2.45 & 7.27 & 3.39 & J130655.8-403519 & -12.46 & 0.98 & uu & 21.42 & 0.0099 \\ \hline 
J1645.8-4533 & -10.36 & 2.56 & 1.30 & 5.04 & J164547.7-453647 & -9.26 & -1.53 & uu & 22.64 & 0.0015 \\ \hline 
J1729.2-3641 & -10.56 & 2.47 & 11.20 & 2.24 & J172916.6-364006 & -12.08 & -0.12 & uu & 21.18 & 0.0085 \\ \hline 
J1838.2+3223 & -11.47 & 2.38 & 2.53 & 2.35 & J183817.0+322416 & -12.15 & 0.29 & uu & 21.30 & 0.0041 \\ \hline 
J1844.4-0306 & -10.26 & 2.60 & 5.72 & 6.52 & J184441.7-030536 & -12.33 & 1.85 & uu & 20.67 & 0.0042 \\ \hline 
J2029.5-4237 & -11.64 & 2.47 & 3.65 & 5.65 & J202913.7-423533 & -11.79 & 0.07 & uu & 21.25 & 0.0022 \\ \hline 
J2039.5-5617 & -10.83 & 2.17 & 13.12 & 8.58 & J203934.3-561706 & -13.00 & 1.33 & uu & 22.33 & 0.0022 \\ \hline 
J2054.2+6904 & -11.38 & 2.38 & 3.78 & 4.40 & J205357.8+690519 & -12.64 & 0.43 & vv & 20.62 & 0.0023 \\ \hline 
 &  &  &  &  &  &  &  &  &  & \\
 &  &  &  &  & (a) \large{$P_{bzr} > 0.99 $} &  &  &  &  & \\
 &  &  &  &  &  \large{\textbf{likely blazars}} &  &  &  &  & \\ \hline
J0004.4-4001 & -11.66 & 2.42 & 18.52 & 0.34 & J000434.1-400036 & -12.41 & 2.50 & uu & 17.93 & 0.9998 \\ \hline
J0017.1-4605 & -11.62 & 2.80 & 4.12 & 0.74 & J001708.5-460612 & -12.46 & 1.72 & uu & 18.92 & 0.9993 \\ \hline 
J0025.4-4838 & -11.69 & 2.12 & 10.13 & 1.30 & J002536.8-483808 & -12.30 & 2.33 & uu & 19.37 & 0.9997 \\ \hline 
J0026.1-0732 & -11.77 & 2.07 & 1.25 & 0.08 & J002611.9-073115 & -11.35 & 2.13 & uu & 19.42 & 0.9998 \\ \hline 
J0031.5-5648 & -11.81 & 1.92 & 11.68 & 1.19 & J003135.1-564640 & -12.39 & 2.43 & w1 & 20.12 & 0.9998 \\ \hline 
J0037.2-2653 & -11.86 & 2.46 & 3.65 & 0.44 & J003729.6-265043 & -12.33 & 1.63 & uu & 20.00 & 0.9996 \\ \hline 
J0057.9+6326 & -11.57 & 1.67 & 11.42 & 2.05 & J005758.1+632642 & -11.56 & 2.04 & vv & 19.11 & 0.9997 \\ \hline 
J0058.3-4603 & -11.94 & 2.10 & 12.61 & 0.64 & J005806.4-460417 & -11.52 & 1.62 & uu & 21.19 & 0.9998 \\ \hline 
J0118.3-6008 & -11.96 & 2.17 & 10.53 & 2.33 & J011824.0-600753 & -12.74 & 2.49 & uu & 19.34 & 0.9994 \\ \hline 
J0120.2-7944 & -11.76 & 2.38 & 15.39 & 1.37 & J011914.7-794510 & -12.32 & 2.09 & uu & 20.13 & 0.9997 \\ \hline 
J0125.9-6303 & -11.78 & 2.32 & 9.33 & 0.10 & J012548.3-630244 & -12.09 & 1.95 & uu & 19.94 & 0.9998 \\ \hline 
J0137.3-3239 & -11.88 & 2.25 & 7.80 & 0.48 & J013724.7-324047 & -12.14 & 2.23 & uu & 21.56 & 0.9998 \\ \hline 
J0140.3+7054 & -11.60 & 2.00 & 6.00 & 1.44 & J014008.6+705250 & -12.36 & 1.99 & uu & 20.64 & 0.9973 \\ \hline 
J0156.3-2420 & -11.58 & 1.86 & 36.97 & 1.81 & J015624.4-242004 & -11.98 & 2.50 & bb & 18.18 & 0.9998 \\ \hline 
J0159.0+3313 & -11.65 & 2.02 & 23.07 & 0.82 & J015905.0+331255 & -12.23 & 2.43 & m2 & 19.43 & 0.9998 \\ \hline 
J0209.8+2626 & -11.69 & 2.25 & 7.14 & 0.52 & J020946.5+262528 & -11.60 & 1.96 & uu & 20.99 & 0.9998 \\ \hline 
J0231.0+3505 & -11.96 & 1.69 & 8.60 & 2.61 & J023112.4+350446 & -11.58 & 1.74 & m2 & 20.39 & 0.9982 \\ \hline 
J0240.2-0248 & -11.58 & 2.55 & 14.69 & 0.12 & J024004.6-024505 & -12.18 & 1.81 & vv & 18.14 & 0.9999 \\ \hline 
J0259.0+0552 & -11.11 & 2.10 & 14.76 & 2.28 & J025857.4+055243 & -12.25 & 1.94 & uu & 18.70 & 0.9991 \\ \hline 
J0301.6-5617 & -11.79 & 1.91 & 20.93 & 1.63 & J030115.2-561648 & -12.57 & 2.29 & uu & 19.49 & 0.9998 \\ \hline 
J0302.5+3354 & -11.49 & 2.32 & 4.87 & 0.70 & J030226.7+335448 & -12.10 & 1.94 & uu & 20.35 & 0.9996 \\ \hline 
J0327.6+2620 & -11.85 & 1.94 & 3.59 & 0.62 & J032737.2+262008 & -12.02 & 2.10 & uu & 20.12 & 0.9997 \\ \hline 
J0333.4-2705 & -11.81 & 2.10 & 9.36 & 0.53 & J033331.5-270918 & -12.01 & 2.12 & m2 & 19.30 & 0.9998 \\ \hline 
J0343.3-6303 & -11.63 & 2.05 & 9.37 & 0.73 & J034323.9-630342 & -12.43 & 3.37 & uu & 20.53 & 0.9998 \\ \hline 
J0406.2+0639 & -11.59 & 2.35 & 14.84 & 0.32 & J040607.7+063919 & -12.48 & 2.07 & w1 & 20.79 & 0.9998 \\ \hline 
J0409.2+2542 & -11.58 & 2.09 & 7.66 & 1.13 & J040921.6+254440 & -12.19 & 2.12 & uu & 19.30 & 0.9997 \\ \hline 
J0427.8-6704 & -11.07 & 2.39 & 13.10 & 0.32 & J042749.5-670435 & -11.81 & -0.14 & uu & 18.14 & 0.9997 \\ \hline 
J0537.5+0959 & -11.29 & 2.62 & 20.40 & 1.30 & J053745.9+095759 & -12.24 & 1.95 & uu & 20.75 & 0.9996 \\ \hline 
J0539.2-6333 & -11.78 & 2.15 & 11.42 & 0.72 & J054002.9-633216 & -12.83 & 2.09 & uu & 18.44 & 0.9998 \\ \hline 
J0553.9-5048 & -11.71 & 2.09 & 8.67 & 1.33 & J055359.2-505150 & -12.21 & 2.01 & uu & 21.11 & 0.9992 \\ \hline 
J0554.2-0259 & -11.45 & 2.42 & 5.83 & 0.47 & J055418.8-025842 & -12.23 & 1.63 & uu & 21.24 & 0.9995 \\ \hline 
J0611.5-2918 & -11.81 & 2.34 & 6.38 & 1.22 & J061141.5-291623 & -12.50 & 2.26 & uu & 21.41 & 0.9987 \\ \hline 
J0620.7-5034 & -12.04 & 1.76 & 20.37 & 2.23 & J062045.7-503350 & -12.26 & 2.37 & uu & 19.93 & 0.9997 \\ \hline 
J0631.0+5626 & -11.79 & 2.11 & 5.26 & 2.17 & J063048.3+562831 & -12.14 & 2.45 & uu & 19.76 & 0.9988 \\ \hline 
J0633.9+5840 & -11.97 & 1.90 & 8.23 & 0.39 & J063400.1+584036 & -12.93 & 2.41 & uu & 21.97 & 0.9998 \\ \hline 
J0650.6+2055 & -11.23 & 1.74 & 7.46 & 1.96 & J065035.4+205557 & -11.20 & 2.16 & vv & 17.05 & 0.9998 \\ \hline 
J0704.3-4829 & -11.51 & 2.09 & 13.09 & 1.64 & J070421.8-482645 & -12.00 & 2.51 & uu & 19.52 & 0.9998 \\ \hline 
J0723.1-3048 & -11.49 & 2.34 & 3.63 & 0.65 & J072310.8-304758 & -11.07 & 1.70 & uu & 20.87 & 0.9996 \\ \hline 
J0725.7-0549 & -11.29 & 1.88 & 5.24 & 2.12 & J072547.5-054830 & -11.61 & 2.38 & uu & 19.15 & 0.9992 \\ \hline 
J0737.4+6535 & -11.82 & 2.05 & 16.09 & 1.47 & J073711.3+653348 & -13.10 & 1.63 & vv & 20.07 & 0.9987 \\ \hline 
J0755.9-0515 & -11.36 & 2.86 & 1.63 & 0.11 & J075614.6-051720 & -11.61 & 1.67 & uu & 20.45 & 0.9998 \\ \hline 
J0800.9+0733 & -11.70 & 1.89 & 3.72 & 2.88 & J080056.5+073235 & -12.28 & 2.59 & vv & 18.54 & 0.9970 \\ \hline 
J0827.0-4111 & -11.45 & 2.12 & 6.90 & 0.51 & J082705.4-411159 & -11.95 & 2.51 & uu & 21.17 & 0.9998 \\ \hline 
J0838.5+4013 & -11.94 & 1.70 & 3.60 & 0.21 & J083902.8+401548 & -12.40 & 2.28 & w2 & 20.38 & 0.9998 \\ \hline 
J0844.9-4117 & -10.85 & 2.68 & 6.33 & 3.20 & J084439.9-411642 & -10.40 & 6.51 & m2 & 14.68 & 0.9999 \\ \hline 
J0858.0-3130 & -11.38 & 1.81 & 12.88 & 0.83 & J085802.9-313046 & -11.59 & 2.48 & vv & 18.32 & 0.9998 \\ \hline 
J0903.5+4057 & -11.94 & 2.04 & 9.30 & 0.32 & J090342.8+405503 & -12.91 & 1.89 & vv & 19.29 & 0.9998 \\ \hline 
J0906.1-1011 & -11.76 & 1.98 & 1.98 & 0.77 & J090616.2-101430 & -12.45 & 1.71 & uu & 20.43 & 0.9933 \\ \hline 
J0910.1-1816 & -11.74 & 1.95 & 8.85 & 2.16 & J091003.9-181613 & -12.01 & 2.30 & w1 & 20.31 & 0.9992 \\ \hline 
J0914.5+6845 & -11.99 & 1.88 & 6.08 & 1.48 & J091430.0+684509 & -12.21 & 2.02 & uu & 20.04 & 0.9993 \\ \hline 
J0928.4-5256 & -11.44 & 1.94 & 10.91 & 0.00 & J092818.7-525701 & -11.59 & 3.16 & uu & 20.17 & 0.9999 \\ \hline 
J0930.9-3030 & -11.71 & 1.78 & 6.73 & 0.95 & J093058.0-303118 & -12.34 & 2.03 & m2 & 19.95 & 0.9996 \\ \hline 
J0934.5+7223 & -11.52 & 2.99 & 10.60 & 0.78 & J093334.0+722101 & -12.38 & 1.55 & uu & 19.15 & 0.9996 \\ \hline 
J0938.8+5155 & -11.91 & 1.98 & 10.82 & 1.07 & J093835.0+515455 & -12.54 & 0.73 & m2 & 20.30 & 0.9963 \\ \hline 
J1008.1-5706 & -10.40 & 2.75 & 3.19 & 3.33 & J100829.1-571242 & -11.05 & 3.96 & uu & 20.21 & 0.9987 \\ \hline 
J1015.5-6030 & -10.73 & 2.79 & 4.76 & 2.36 & J101545.8-602939 & -11.44 & 3.23 & uu & 20.78 & 0.9984 \\ \hline 
J1016.1-4247 & -11.40 & 1.81 & 9.64 & 2.79 & J101620.8-424723 & -11.97 & 2.54 & uu & 18.57 & 0.9995 \\ \hline 
J1016.2-5729 & -10.86 & 2.51 & 9.66 & 1.33 & J101625.7-572807 & -11.23 & 1.35 & uu & 18.39 & 0.9995 \\ \hline 
J1018.1-2705 & -11.47 & 2.52 & 7.02 & 2.27 & J101750.2-270550 & -11.76 & 1.61 & uu & 18.80 & 0.9974 \\ \hline 
J1024.5-4543 & -11.55 & 1.89 & 8.16 & 0.96 & J102432.6-454428 & -11.50 & 2.41 & uu & 19.33 & 0.9998 \\ \hline 
J1024.5-5329 & -11.85 & 1.74 & 9.88 & 3.24 & J102428.3-532806 & -11.40 & 2.30 & w2 & 21.22 & 0.9988 \\ \hline 
J1034.7-4645 & -11.73 & 2.13 & 9.92 & 0.52 & J103438.7-464405 & -11.49 & 2.03 & w1 & 21.31 & 0.9998 \\ \hline 
J1047.2+6740 & -12.03 & 1.96 & 17.73 & 0.53 & J104705.6+673806 & -12.34 & 2.02 & w1 & 21.34 & 0.9998 \\ \hline 
J1048.4-5030 & -11.83 & 1.66 & 5.91 & 1.27 & J104824.2-502941 & -12.52 & 3.02 & bb & 18.93 & 0.9998 \\ \hline 
J1049.8+2741 & -11.77 & 2.00 & 11.34 & 0.65 & J104938.8+274217 & -12.46 & 2.60 & bb & 18.65 & 0.9998 \\ \hline 
J1058.4-6625 & -11.33 & 1.93 & 8.91 & 1.76 & J105833.1-662607 & -12.09 & 1.77 & uu & 21.38 & 0.9921 \\ \hline 
J1106.7+3623 & -11.65 & 2.79 & 30.30 & 0.61 & J110636.7+362648 & -12.50 & 1.06 & uu & 19.31 & 0.9998 \\ \hline 
J1111.4+0137 & -11.78 & 2.28 & 2.75 & 0.29 & J111114.2+013431 & -11.66 & 1.92 & uu & 20.51 & 0.9998 \\ \hline 
J1114.6+1225 & -11.78 & 2.30 & 6.55 & 2.20 & J111436.7+122718 & -12.64 & 2.54 & uu & 21.06 & 0.9944 \\ \hline 
J1119.9-1007 & -11.69 & 2.19 & 18.56 & 1.83 & J111948.2-100704 & -12.38 & 1.88 & uu & 19.52 & 0.9996 \\ \hline 
J1122.0-0231 & -11.59 & 2.44 & 4.75 & 1.60 & J112213.8-022916 & -12.00 & 2.41 & uu & 18.72 & 0.9995 \\ \hline 
J1123.8-4552 & -11.34 & 2.21 & 8.29 & 0.81 & J112355.5-455019 & -12.28 & 1.21 & w1 & 19.01 & 0.9994 \\ \hline 
J1146.0-0638 & -11.46 & 1.75 & 5.11 & 1.44 & J114600.8-063851 & -12.00 & 2.13 & uu & 19.70 & 0.9992 \\ \hline 
J1155.2-1111 & -11.75 & 1.92 & 9.12 & 1.42 & J115514.7-111125 & -12.29 & 1.80 & uu & 21.14 & 0.9985 \\ \hline 
J1220.1-2458 & -11.58 & 1.88 & 8.77 & 0.79 & J122014.5-245949 & -11.54 & 1.92 & uu & 19.86 & 0.9998 \\ \hline 
J1224.6+7011 & -11.41 & 3.01 & 9.72 & 1.00 & J122457.2+700729 & -12.47 & 2.07 & uu & 16.06 & 0.9998 \\ \hline 
J1239.7-3455 & -11.61 & 2.67 & 10.37 & 0.00 & J123934.1-345429 & -12.06 & 1.72 & uu & 21.52 & 0.9999 \\ \hline 
J1243.7+1727 & -11.92 & 1.87 & 1.50 & 1.01 & J124351.6+172643 & -12.94 & 2.86 & uu & 19.72 & 0.9983 \\ \hline 
J1250.9-4943 & -11.22 & 2.11 & 20.98 & 2.07 & J125058.5-494444 & -12.53 & 2.35 & w2 & 21.98 & 0.9983 \\ \hline 
J1256.8+5329 & -11.35 & 2.64 & 5.01 & 0.88 & J125630.4+533203 & -12.55 & 1.66 & uu & 19.19 & 0.9985 \\ \hline 
J1320.3-6410 & -11.24 & 2.60 & 14.06 & 3.74 & J132016.9-641353 & -9.00 & 8.19 & uu & 17.93 & 0.9999 \\ \hline 
J1357.3-6123 & -10.77 & 2.62 & 7.88 & 5.37 & J135657.9-612313 & -11.14 & 3.63 & m2 & 20.75 & 0.9985 \\ \hline 
J1401.7-3217 & -11.61 & 2.14 & 27.08 & 1.31 & J140200.0-321844 & -12.51 & 2.26 & uu & 20.52 & 0.9998 \\ \hline 
J1415.9-1504 & -11.68 & 2.12 & 11.09 & 1.46 & J141546.1-150229 & -12.23 & 2.48 & w1 & 19.50 & 0.9997 \\ \hline 
J1429.8-0739 & -11.95 & 2.31 & 3.32 & 3.04 & J142949.7-073302 & -12.14 & 1.87 & uu & 17.58 & 0.9901 \\ \hline 
J1514.8+4448 & -11.13 & 2.36 & 56.34 & 3.43 & J151451.0+444957 & -12.84 & 1.42 & vv & 19.22 & 0.9969 \\ \hline 
J1528.4+2004 & -11.90 & 2.02 & 7.88 & 0.28 & J152836.0+200424 & -12.08 & 1.94 & uu & 20.80 & 0.9998 \\ \hline 
J1545.0-6642 & -11.37 & 1.74 & 2.78 & 0.18 & J154459.0-664148 & -11.01 & 2.11 & uu & 18.37 & 0.9998 \\ \hline 
J1557.2+3822 & -12.05 & 1.95 & 8.29 & 1.51 & J155711.9+382032 & -12.54 & 2.27 & uu & 21.45 & 0.9990 \\ \hline 
J1623.7-2315 & -11.33 & 2.37 & 7.72 & 0.79 & J162334.1-231750 & -11.72 & 1.13 & uu & 18.57 & 0.9996 \\ \hline 
J1631.8+4144 & -11.87 & 1.78 & 11.29 & 0.79 & J163146.7+414634 & -12.37 & 2.45 & uu & 19.31 & 0.9998 \\ \hline 
J1639.8-4642 & -10.24 & 2.51 & 3.14 & 3.83 & J163946.2-464058 & -9.48 & 8.73 & uu & 17.46 & 0.9999 \\ \hline 
J1644.8+1850 & -11.61 & 1.99 & 9.12 & 2.37 & J164457.3+185149 & -12.50 & 2.26 & uu & 19.34 & 0.9986 \\ \hline 
J1645.0+1654 & -11.66 & 2.21 & 8.37 & 0.76 & J164500.0+165510 & -12.14 & 2.34 & uu & 19.57 & 0.9998 \\ \hline 
J1648.7+4834 & -11.89 & 2.14 & 8.53 & 1.00 & J164900.4+483417 & -12.33 & 2.28 & uu & 19.60 & 0.9998 \\ \hline 
J1650.9-4420 & -10.59 & 2.59 & 4.96 & 4.18 & J165124.2-442142 & -7.39 & 10.00 & m2 & 20.23 & 0.9999 \\ \hline 
J1651.7-7241 & -11.87 & 2.04 & 2.03 & 0.40 & J165151.5-724310 & -11.98 & 1.17 & uu & 20.64 & 0.9992 \\ \hline 
J1706.5-4023 & -10.54 & 2.55 & 13.11 & 2.85 & J170633.3-402544 & -8.46 & 9.03 & w1 & 18.48 & 0.9999 \\ \hline 
J1719.1-5348 & -11.44 & 2.16 & 6.11 & 0.12 & J171856.6-535042 & -11.22 & 1.76 & vv & 19.46 & 0.9998 \\ \hline 
J1720.6-5144 & -11.54 & 2.16 & 7.02 & 1.47 & J172032.7-514413 & -11.14 & 1.87 & uu & 19.70 & 0.9997 \\ \hline 
J1752.0+3606 & -11.90 & 1.95 & 10.88 & 1.95 & J175209.6+360632 & -12.20 & 1.76 & uu & 19.55 & 0.9992 \\ \hline 
J1817.2-3035 & -11.05 & 2.26 & 9.39 & 1.32 & J181720.4-303257 & -11.71 & 1.80 & w1 & 21.75 & 0.9979 \\ \hline 
J1818.5+2533 & -11.36 & 2.67 & 9.96 & 1.79 & J181830.9+253707 & -12.62 & 1.25 & uu & 16.99 & 0.9980 \\ \hline 
J1820.4-1609 & -10.25 & 2.61 & 10.37 & 0.78 & J182029.6-161044 & -11.08 & 1.94 & bb & 16.66 & 0.9998 \\ \hline 
J1848.7-6307 & -11.98 & 1.95 & 5.40 & 0.27 & J184838.3-630544 & -12.49 & 1.84 & uu & 20.06 & 0.9998 \\ \hline 
J1856.1-1222 & -11.19 & 2.18 & 10.50 & 1.17 & J185606.7-122149 & -11.79 & 1.92 & uu & 20.74 & 0.9995 \\ \hline 
J1918.0+0331 & -11.44 & 1.79 & 14.45 & 1.38 & J191803.6+033030 & -11.91 & 2.46 & w1 & 21.08 & 0.9997 \\ \hline 
J1927.5+0154 & -11.50 & 1.89 & 16.83 & 1.16 & J192731.3+015357 & -11.86 & 2.35 & uu & 19.78 & 0.9998 \\ \hline 
J1955.3-5032 & -11.59 & 2.47 & 13.03 & 1.22 & J195512.5-503012 & -11.98 & 2.14 & uu & 20.33 & 0.9997 \\ \hline 
J2008.4+1619 & -11.60 & 2.10 & 8.76 & 1.73 & J200827.6+161844 & -11.83 & 2.08 & uu & 20.49 & 0.9993 \\ \hline 
J2027.0+3343 & -11.50 & 1.77 & 2.96 & 1.04 & J202658.7+334303 & -10.96 & 1.56 & uu & 19.94 & 0.9994 \\ \hline 
J2030.5+2235 & -11.75 & 1.78 & 9.27 & 2.45 & J203031.2+223435 & -12.42 & 2.10 & uu & 21.11 & 0.9918 \\ \hline 
J2041.1-6138 & -11.58 & 2.21 & 7.30 & 0.87 & J204112.0-613952 & -12.34 & 2.17 & uu & 18.98 & 0.9997 \\ \hline 
J2046.9-5409 & -11.77 & 2.01 & 9.84 & 1.07 & J204700.5-541246 & -12.38 & 1.85 & w1 & 20.08 & 0.9996 \\ \hline 
J2058.5-1833 & -12.09 & 1.25 & 7.64 & 0.59 & J205836.5-183102 & -12.62 & 1.17 & vv & 19.51 & 0.9997 \\ \hline 
J2109.6+3954 & -11.42 & 1.61 & 11.90 & 3.23 & J210936.4+395513 & -11.85 & 3.29 & uu & 21.16 & 0.9995 \\ \hline 
J2114.9-3326 & -11.48 & 2.25 & 8.92 & 0.61 & J211452.0-332532 & -11.69 & 2.24 & uu & 19.30 & 0.9998 \\ \hline 
J2159.6-4620 & -11.54 & 1.68 & 13.82 & 0.80 & J215935.9-461954 & -12.34 & 2.72 & uu & 18.82 & 0.9998 \\ \hline 
J2207.1+2222 & -11.89 & 1.87 & 14.48 & 0.06 & J220704.3+222234 & -12.45 & 2.52 & uu & 19.91 & 0.9999 \\ \hline 
J2222.9+1507 & -11.91 & 1.91 & 6.46 & 1.86 & J222253.9+151052 & -12.58 & 2.24 & uu & 18.88 & 0.9992 \\ \hline 
J2225.8-0804 & -11.67 & 1.99 & 11.40 & 0.80 & J222552.9-080416 & -12.42 & 1.83 & uu & 20.56 & 0.9997 \\ \hline 
J2237.2-6726 & -11.72 & 1.99 & 10.90 & 0.66 & J223709.3-672614 & -12.01 & 1.90 & uu & 20.32 & 0.9998 \\ \hline 
J2240.3-5241 & -11.24 & 2.07 & 15.19 & 1.09 & J224017.5-524117 & -12.71 & 1.84 & uu & 18.54 & 0.9997 \\ \hline 
J2247.7-5857 & -11.68 & 2.57 & 6.07 & 0.53 & J224745.0-585501 & -12.31 & 2.61 & uu & 20.60 & 0.9998 \\ \hline 
J2303.9+5554 & -11.65 & 1.78 & 9.08 & 1.62 & J230351.7+555618 & -11.50 & 1.93 & uu & 20.76 & 0.9995 \\ \hline 
J2311.6-4427 & -11.83 & 2.24 & 8.93 & 1.31 & J231145.6-443221 & -12.72 & 2.76 & uu & 19.33 & 0.9997 \\ \hline 
J2317.7+2839 & -11.48 & 1.96 & 20.83 & 0.64 & J231740.0+283954 & -12.94 & 2.30 & vv & 19.44 & 0.9998 \\ \hline 
J2326.9-4130 & -11.55 & 2.73 & 10.51 & 0.61 & J232653.2-412713 & -12.44 & 2.00 & m2 & 17.25 & 0.9998 \\ \hline 
J2331.6+4430 & -11.91 & 1.98 & 11.81 & 0.43 & J233129.6+443101 & -12.49 & 2.04 & uu & 21.30 & 0.9998 \\ \hline 
J2336.9-8427 & -11.75 & 2.01 & 8.00 & 0.00 & J233627.1-842648 & -12.11 & 2.02 & uu & 20.06 & 0.9999 \\ \hline 
J2337.7-2903 & -11.92 & 2.15 & 3.20 & 1.82 & J233730.2-290241 & -12.06 & 2.42 & uu & 19.97 & 0.9982 \\ \hline 
J2347.9-5106 & -11.08 & 2.53 & 59.61 & 0.42 & J234804.2-510748 & -12.54 & 0.73 & uu & 22.29 & 0.9997 \\ \hline 
 &  &  &  &  &  &  &  &  &  & \\
 &  &  &  &  & (c) \large{$0.01 < P_{bzr} < 0.99$} &  &  &  &  & \\
 &  &  &  &  &  \large{\textbf{ambiguous}} &  &  &  &  & \\ \hline
J0251.1-1830 & -11.71 & 1.62 & 5.06 & 3.70 & J025111.4-183114 & -12.20 & 1.98 & uu & 20.18 & 0.7902 \\ \hline 
J0341.9+3153 & -11.05 & 2.47 & 8.89 & 1.49 & J034158.1+314850 & -11.75 & 0.33 & uu & 21.14 & 0.5278 \\ \hline 
J0407.7-5702 & -11.80 & 2.41 & 3.32 & 1.47 & J040731.2-570022 & -12.06 & 1.50 & uu & 21.65 & 0.8723 \\ \hline 
J0502.2+3016 & -11.74 & 1.96 & 6.27 & 2.38 & J050212.7+301934 & -11.56 & 1.44 & uu & 20.59 & 0.9854 \\ \hline 
J0540.7+3611 & -11.34 & 2.55 & 6.84 & 4.10 & J054046.9+361657 & -11.86 & 1.31 & uu & 20.63 & 0.1307 \\ \hline 
J0622.5+3120 & -11.60 & 2.60 & 9.98 & 3.05 & J062254.0+312037 & -11.94 & 1.50 & uu & 20.34 & 0.9621 \\ \hline 
J0624.7-4903 & -11.64 & 2.62 & 3.95 & 1.50 & J062419.5-490639 & -12.55 & 2.04 & m2 & 20.59 & 0.9813 \\ \hline 
J0700.2-5118 & -11.73 & 2.16 & 1.23 & 1.62 & J070033.8-512011 & -11.84 & 1.86 & w2 & 19.52 & 0.9215 \\ \hline 
J0838.7-2827 & -11.06 & 2.12 & 5.84 & 6.48 & J083843.4-282701 & -11.29 & 1.69 & w1 & 17.40 & 0.8690 \\ \hline 
J0859.2-4729 & -10.71 & 2.54 & 6.94 & 3.88 & J085905.7-473041 & -11.41 & 1.72 & uu & 19.86 & 0.7309 \\ \hline 
J1126.0-5007 & -11.83 & 2.13 & 5.98 & 4.58 & J112624.9-500808 & -11.84 & 1.67 & vv & 18.29 & 0.9657 \\ \hline 
J1312.6-6231 & -11.03 & 2.43 & 2.16 & 3.54 & J131230.7-623430 & -11.95 & 1.14 & uu & 17.57 & 0.0449 \\ \hline 
J1410.7+7405 & -11.42 & 1.86 & 9.17 & 4.15 & J141044.9+740511 & -12.83 & 2.30 & uu & 18.74 & 0.9651 \\ \hline 
J1441.4-1934 & -11.69 & 1.80 & 3.76 & 3.27 & J144127.7-193549 & -12.06 & 2.26 & uu & 19.67 & 0.9632 \\ \hline 
J1619.4-5106 & -10.60 & 2.33 & 1.71 & 2.12 & J161935.6-510613 & -12.05 & 2.45 & w1 & 20.39 & 0.3222 \\ \hline 
J1627.7+3219 & -11.47 & 2.15 & 6.98 & 6.60 & J162743.0+322102 & -12.47 & 2.39 & uu & 17.42 & 0.9413 \\ \hline 
J1700.2-4237 & -10.59 & 2.74 & 5.15 & 4.26 & J165955.9-423023 & -12.20 & 1.58 & uu & 19.91 & 0.0205 \\ \hline 
J1701.3-4924 & -11.20 & 2.48 & 5.55 & 2.51 & J170136.3-492402 & -12.18 & 1.55 & uu & 20.48 & 0.5348 \\ \hline 
J1733.4+2235 & -11.90 & 2.16 & 4.61 & 4.06 & J173330.6+223615 & -12.58 & 1.94 & uu & 19.51 & 0.5335 \\ \hline 
J1734.0-2933 & -11.11 & 2.39 & 4.99 & 3.74 & J173411.0-293117 & -12.08 & 2.06 & w1 & 20.91 & 0.2448 \\ \hline 
J1750.8-3106 & -11.45 & 2.38 & 5.82 & 3.76 & J175052.3-310255 & -12.40 & 1.45 & vv & 19.40 & 0.2285 \\ \hline 
J1836.9+4439 & -11.93 & 1.82 & 8.62 & 1.59 & J183703.6+443812 & -12.13 & 0.78 & w1 & 21.18 & 0.9567 \\ \hline 
J1904.7-0708 & -10.88 & 2.46 & 17.00 & 6.85 & J190444.5-070743 & -12.23 & 2.23 & uu & 20.36 & 0.5493 \\ \hline 
J1948.9+3414 & -11.60 & 2.28 & 11.41 & 4.39 & J194916.8+341050 & -11.82 & 1.40 & uu & 18.88 & 0.9720 \\ \hline 
J2012.1-5234 & -11.75 & 1.80 & 2.94 & 2.26 & J201213.8-523246 & -11.86 & 1.80 & uu & 19.60 & 0.9723 \\ \hline 
J2021.9+3609 & -11.66 & 1.57 & 6.09 & 2.80 & J202201.1+361106 & -12.17 & 1.98 & uu & 20.77 & 0.9538 \\ \hline 
J2056.4+4351 & -10.84 & 2.64 & 2.77 & 3.55 & J205551.3+435220 & -10.74 & 2.59 & m2 & 21.05 & 0.9364 \\ \hline 
J2105.9+7508 & -11.42 & 2.22 & 2.47 & 1.26 & J210606.8+750924 & -12.38 & 2.02 & uu & 21.49 & 0.8837 \\ \hline 
\enddata
\end{deluxetable*}

\end{longrotatetable}
\begin{longrotatetable}
\begin{deluxetable*}{ccccccccccc}
\tablecaption{\textit{Fermi}-LAT features for the unassociated sample investigated in this work, along with \textit{Swift}-XRT and -UVOT parameters for all possible X-ray/UV/optical counterparts. Only unassociated sources with multiple notable X-ray sources within the \textit{Fermi}-LAT uncertainly ellipse are included here. The extraction and derivation of the parameters here are described in sections \ref{sec:Analysis} and \ref{sec:Samples}. The $m_V$ estimates were produced using the noted UVOT filter, the closest available to the V band central wavelength. Two sources, noted with *asterisks*, are coincident with dim catalogued stars. UV/optical emission from those stars is probably not related to a pulsar or blazar, and thus the $P_{bzr}$ values should not be trusted. \label{tab:xrt2}}

\tablewidth{\columnwidth}
\tablehead{
\colhead{Target} & \colhead{$\log(F_G)$} &\colhead{$\Gamma_G$} &
\colhead{Vari. Ind.} & \colhead{Curv. Ind.}&\colhead{XRT source}& \colhead{$\log(F_X)$} & \colhead{$\Gamma_X$} & \colhead{UVOT Filter} & \colhead{$m_V$} & \colhead{$P_{bzr}$} \\
\colhead{4FGL} & \colhead{$\log(\rm{erg/s/cm^2})$} &\colhead{} &
\colhead{} & \colhead{} & \colhead{SwXF4} & \colhead{$\log(\rm{erg/s/cm^2})$} & \colhead{} & \colhead{} & \colhead{} & \colhead{}
}
\startdata
J0126.3-6746 & -12.09 & 1.74 & 4.00 & 1.22 & J012612.0-674746 & -12.38 & 2.54 & uu & 21.64 & 0.9991 \\ 
 &  &  &  &  & J012621.6-674638 & -12.81 & 2.02 & uu & 20.56 & 0.9968 \\ \arrayrulecolor{gray} \hline 
J0523.3-2527 & -10.93 & 2.08 & 1.65 & 10.68 & J052317.1-252737 & -12.27 & 1.46 & vv & 16.86 & 0.0026 \\ 
 &  &  &  &  & J052323.9-252737 & -13.33 & 1.50 & vv & 20.70 & 0.0015 \\ \hline 
J0544.8+5209 & -11.55 & 2.55 & 7.56 & 0.55 & J054424.5+521513 & -12.28 & 1.03 & uu & 20.86 & 0.9991 \\ 
 &  &  &  &  & J054456.7+520847 & -12.55 & 2.53 & uu & 20.87 & 0.9998 \\ \hline 
J0610.8-4911 & -11.83 & 2.01 & 6.58 & 0.40 & J061031.8-491222 & -12.73 & 2.58 & uu & 13.65 & 0.9999 \\ 
 &  &  &  &  & J061100.0-491034 & -11.67 & 2.01 & uu & 20.78 & 0.9998 \\ \hline 
J0738.6+1311 & -11.54 & 2.53 & 10.49 & 0.78 & *J073843.4+131330* & -12.54 & 2.29 & uu & 13.37 & 0.9998 \\ 
 &  &  &  &  & J073848.7+130755 & -12.72 & 2.48 & uu & 21.01 & 0.9997 \\ \hline 
J0800.1-5531 & -11.53 & 2.54 & 6.63 & 0.05 & J075949.3-553254 & -12.58 & 1.67 & w1 & 18.38 & 0.9999 \\ 
 &  &  &  &  & J080013.1-553408 & -12.53 & 1.12 & w1 & 19.85 & 0.9998 \\ \hline 
J1011.1-4420 & -11.72 & 2.05 & 5.42 & 0.32 & J101124.5-441856 & -12.72 & 1.79 & vv & 19.17 & 0.9998 \\ 
 &  &  &  &  & J101132.0-442255 & -10.95 & 1.71 & vv & 18.21 & 0.9998 \\ \hline 
J1018.1-4051 & -11.56 & 2.53 & 9.92 & 1.63 & J101801.5-405520 & -12.60 & 2.00 & m2 & 19.39 & 0.9990 \\  
 &  &  &  &  & J101807.6-404408 & -12.66 & 2.07 & uu & 21.02 & 0.9966 \\ \hline 
J1326.0+3507 & -11.96 & 2.24 & 27.19 & 1.33 & J132544.4+350450 & -12.52 & 1.31 & uu & 19.78 & 0.9997 \\ 
 &  &  &  &  & J132622.3+350628 & -12.86 & 1.98 & uu & 19.51 & 0.9998 \\ \hline 
J1637.5+3005 & -11.68 & 2.71 & 7.80 & 0.54 & J163728.2+300958 & -12.95 & 1.92 & uu & 19.80 & 0.9997 \\ 
 &  &  &  &  & J163738.4+300513 & -12.68 & 2.00 & uu & 19.80 & 0.9997 \\ 
 &  &  &  &  & J163739.3+301015 & -11.91 & 0.71 & uu & 19.69 & 0.9995 \\ \hline 
J1808.5-3701 & -11.37 & 2.39 & 8.88 & 4.05 & J180822.8-370421 & -14.08 & 0.57 & vv & 21.67 & 0.0019 \\ 
 &  &  &  &  & J180827.6-365842 & -10.89 & 1.54 & vv & 19.49 & 0.9936 \\ 
 &  &  &  &  & J180834.9-370211 & -13.72 & 0.01 & vv & 18.15 & 0.0032 \\  
 &  &  &  &  & J180841.8-370220 & -13.40 & 1.46 & vv & 20.22 & 0.0177 \\  \hline
J1846.9-0227 & -10.47 & 2.47 & 6.89 & 5.24 & *J184650.7-022904* & -8.97 & 7.90 & uu & 13.96 & 0.9999 \\ 
 &  &  &  &  & J184651.6-022507 & -11.62 & 4.12 & uu & 21.23 & 0.9966 \\ \hline 
J1910.8+2856 & -11.39 & 1.83 & 19.18 & 1.39 & J191052.2+285624 & -11.47 & 2.27 & w2 & 20.60 & 0.9998 \\ 
 &  &  &  &  & J191059.4+285635 & -12.41 & 1.48 & w2 & 19.64 & 0.9995 \\ \hline 
J2351.4-2818 & -11.79 & 2.37 & 1.82 & 1.04 & J235136.5-282154 & -12.52 & 2.27 & uu & 16.66 & 0.9995 \\ 
 &  &  &  &  & J235138.1-281823 & -12.44 & 2.06 & uu & 21.46 & 0.9547 \\ \hline 
\enddata
\end{deluxetable*}
\end{longrotatetable}
\begin{deluxetable*}{ccc}
\tablecaption{Results from a SIMBAD position cross-reference for the possible counterparts to unassociated sources with multiple notable excesses in the gamma-ray uncertainty ellipse, helping discriminate excesses that might be eliminated from consideration as possible counterparts with additional astrophysical information or description. \label{tab:xrtmultcross}}

\tablewidth{\columnwidth}
\tablehead{
\colhead{Target name} & \colhead{XRT excess name} & \colhead{Cross-reference results} \\
\colhead{4FGL} & \colhead{SwXF4} & \colhead{}  }
\startdata
J0126.3-6746 & J012612.0-674746 &  \\ 
 & J012621.6-674638 &  \\ 
\hline J0523.3-2527 & J052317.1-252737 & 2.3" from  TIC 31091702, possible RR Lyrae star \\ 
 & J052323.9-252737 &  \\ 
\hline J0544.8+5209 & J054424.5+521513 &  \\ 
 & J054456.7+520847 &  \\ 
\hline J0610.8-4911 & J061031.8-491222 & 0.6" from  UCAC3 82-17607, high-proper-motion star  \\ 
 & J061100.0-491034 & 2.0" from  GALEX J061100.1-491033, quasar \\ 
\hline J0738.6+1311 & J073843.4+131330 & 4.2" from  TYC 777-755-1, star  \\ 
 & J073848.7+130755 &  \\ 
\hline J0800.1-5531 & J075949.3-553254 &  \\ 
 & J080013.1-553408 &  \\ 
\hline J1011.1-4420 & J101124.5-441856 &  \\ 
 & J101132.0-442255 &  \\ 
\hline J1018.1-4051 & J101801.5-405520 &  \\ 
 & J101807.6-404408 &  \\ 
\hline J1326.0+3507 & J132544.4+350450 & 5.3" from  LSPM J1325+3504, high-proper-motion star  \\ 
 & J132622.3+350628 & 3.7" from  LAMOST J132622.19+350624.4, quasar  \\ 
\hline J1637.5+3005 & J163728.2+300958 &  \\ 
 & J163738.4+300513 & 5.9" from  BWE 1635+3010, BL Lac  \\ 
 & J163739.3+301015 & 2.4" from  2MASS J16373918+3010130, Seyfert 1 galaxy  \\ 
\hline J1808.5-3701 & J180822.8-370421 &  \\ 
 & J180827.6-365842 & 2.0" from V* V4580 Sgr, low-mass X-ray binary  \\ 
 & J180834.9-370211 &  \\ 
 & J180841.8-370220 &  \\ 
\hline J1846.9-0227 & J184650.7-022904 & 3.3" from  BD-02 4739, star  \\ 
 & J184651.6-022507 &  \\ 
\hline J1910.8+2856 & J191052.2+285624 & Recently categorized as BL Lac in 4FGL lists  \\ 
 & J191059.4+285635 & 3.7" from V* V584 Lyr, cataclysmic variable star  \\ 
\hline J2351.4-2818 & J235136.5-282154 & 3.18" from  IC 5362, bright cluster galaxy  \\ 
 & J235138.1-281823 &  \\ 
\enddata
\end{deluxetable*}

\end{document}